\documentclass[10pt,journal,compsoc]{IEEEtran}

\ifCLASSOPTIONcompsoc
  \usepackage[nocompress]{cite}
\else
  \usepackage{cite}
\fi
\usepackage{hyperref}
\usepackage{array}
\usepackage{float,epsfig,enumerate,bm,amsmath,color,soul,amssymb}
\usepackage{algorithm}
\usepackage{algorithmic}
\usepackage{graphicx}

\usepackage[labelformat=simple]{subcaption}

\usepackage{caption}
\newcommand{\minew}[1]{{\color{black}{#1}}}

\usepackage{ulem}

\usepackage{amsthm}
\theoremstyle{plain}
\newtheorem{theorem}{Proposition}

\ifCLASSINFOpdf
\else
\fi

\begin{document}

\title{A Wireless AI-Generated Content (AIGC) Provisioning Framework Empowered by Semantic Communication}

\author{Runze~Cheng,~\IEEEmembership{Member,~IEEE,}
        Yao~Sun,~\IEEEmembership{Senior~Member,~IEEE,}
        Dusit Niyato,~\IEEEmembership{Fellow,~IEEE,}
        Lan Zhang,~\IEEEmembership{Member,~IEEE,}
        Lei Zhang,~\IEEEmembership{Senior~Member,~IEEE,}
        and~Muhammad~Imran,~\IEEEmembership{Fellow,~IEEE}

\thanks{Runze Cheng, Yao Sun, Lei Zhang, and Muhammad Ali Imran are with the James Watt School of Engineering, University of Glasgow, Glasgow G12 8QQ, U.K.}
\thanks{Dusit Niyato is with the School of Computer Science and Engineering, Nanyang Technological University, 639798, Singapore.}
\thanks{Lan Zhang is with the Department of Electrical and Computer Engineering, Clemson University, 29634, South Carolina, United States.}
\thanks{The corresponding code can be found at \url{https://github.com/RoyChengsCode/SemAIGCtransceiver.git}.}
\thanks{Yao Sun is the corresponding author. (Email: Yao.Sun@glasgow.ac.uk)}
}

\IEEEtitleabstractindextext{%
\begin{abstract}
    \minew{
    With the significant advances in AI-generated content (AIGC) and the proliferation of mobile devices, providing high-quality AIGC services via wireless networks is becoming the future direction. 
    However, the primary challenges of AIGC services provisioning in wireless networks lie in unstable channels, limited bandwidth resources, and unevenly distributed computational resources.
    To this end, this paper proposes a semantic communication (SemCom)-empowered AIGC (SemAIGC) generation and transmission framework, where only semantic information of the content rather than all the binary bits should be generated and transmitted by using SemCom.
    Specifically, SemAIGC integrates diffusion models within the semantic encoder and decoder to design a workload-adjustable transceiver thereby allowing adjustment of computational resource utilization in edge and local. 
    In addition, a \uline{r}esource-aware w\uline{o}rkl\uline{o}ad \uline{t}rade-off (ROOT) scheme is devised to intelligently make workload adaptation decisions for the transceiver, thus efficiently generating, transmitting, and fine-tuning content as per dynamic wireless channel conditions and service requirements.
    Simulations verify the superiority of our proposed SemAIGC framework in terms of latency and content quality compared to conventional approaches. 
    }
\end{abstract}

}

\maketitle
\IEEEdisplaynontitleabstractindextext

\IEEEpeerreviewmaketitle

\IEEEraisesectionheading{\section{Introduction}\label{sec:introduction}}

\IEEEPARstart{G}{enerative} AI has demonstrated remarkable strides in analyzing various forms of media and creating AI-generated content (AIGC) \cite{jo2023promise}. 
For the delivery of high-quality content, well-crafted AIGC models have evolved to maintain millions of parameters within their neural networks. 
This feat can necessitate a large-scale server devouring thousands of hours of graphics processing unit (GPU) time in a single model training session \cite{wang2023survey}. 
Limited by the computing-intensive nature of AIGC, existing AIGC applications, including notable products like ChatGPT, DELL-R-2, and ERNIE Bot, predominantly find their home on cloud servers equipped with ample computational resources \cite{roumeliotis2023chatgpt}. 

\subsection{Background: AIGC in Wireless}

Driven by the ubiquitous wireless connectivity and surge global traffic of terminal devices \cite{penalvo2022mobile}, many AI companies are actively strategizing and positioning themselves in the realm of wireless device-oriented AIGC applications. 
However, wireless AIGC services demand strict requirements on content quality and latency, which pose significant challenges under dynamic wireless channels. 
Moreover, these services often grapple with limited communication resources (like bandwidth, and high-quality channels) and unevenly distributed computational resources in wireless networks. 
Therefore, it calls urgently for efficient and accessible AIGC generation and transmission frameworks underpinned by wireless networks. 

Recently, a few studies about wireless AIGC generation and transmission have been conducted to improve the utilization of unevenly distributed computational resources, as listed in Table \ref{tab1}. 
The authors in \cite{xu2023unleashing} propose a wireless AIGC network architecture where cloud servers primarily handle computationally intensive AIGC model pre-training. 
Additionally, it offloads content generation tasks from cloud servers to edge transmitters.
In \cite{du2023enabling}, a dynamic AIGC service provider selection scheme is proposed in an intelligent wireless AIGC network to enable users to connect to the provider with suitable edge servers and enough computational resources. 
Considering the limited bandwidth resources and dynamic channel quality, the authors in \cite{wang2023unified} develop a pricing-based incentive mechanism for AIGC generation and transmission framework to maximize users' utility in mobile edge networks.    
These existing works didn't address the insufficient communication resources in wireless networks when coping with the potentially overwhelming number of service requests.

In parallel, some research works aim to deploy AIGC models in local devices to address the communication resource shortage in the wireless network, as listed in Table \ref{tab1}. 
\minew{Lightweight AIGC models have been proposed and tested in works \cite{park2022real,sun2020mobilebert}.
Moreover, some experiments validate the capability of mobile devices for AIGC model deployment, for example, an image can be generated within 2 seconds on a typical smartphone \cite{li2024snapfusion}.}
Under this premise, a novel collaborative distributed AIGC framework is conceptualized in \cite{du2023exploring}, which allows smartphones to process the diffusion model locally in an edge server-empowered wireless AIGC network. 
In this way, less content is transmitted from edge to local, thus the pressure of communication resources is relieved.
However, considering the computational resource limitation of mobile devices, the generated content quality and content-generating latency are difficult to ensure. 

\begin{table}[h]
    \centering
    \caption{\minew{Comparison of different AIGC frameworks}}
    \begin{tabular}{|m{1.5cm}|m{4.5cm}|m{1.3cm}|}
        \hline
        \textbf{AIGC Frameworks} & \textbf{Features} & \textbf{Related Literatures} \\ \hline
        Conventional Cloud AIGC & These models run on remote servers, generally with high content quality and low computation latency, but lead to high transmission latency, and high core network access latency & \cite{jo2023promise},\cite{wang2023survey},\cite{roumeliotis2023chatgpt}\\ \hline
        Edge and Local AIGC & Deploying AIGC models in edge servers, and local devices, flexibly choosing suitable edge servers or local devices to generate contents as per communication resource and computational ability.  & \cite{xu2023unleashing},\cite{du2023enabling},\cite{wang2023unified},\cite{park2022real},\cite{sun2020mobilebert},\cite{li2024snapfusion},\cite{du2023exploring} \\ \hline
        SemCom-empowered AIGC & Integrating SemCom into the AIGC model and deploying modularized components at both the transmitter and receiver, thus enabling the transmission of semantic information between them, and allowing edge servers and local devices to collaborate in content generation using distributed computational resources. & \cite{liu2024semantic} \\ \hline    
    \end{tabular}
    \label{tab1}
\end{table}

\subsection{Motivation}
\minew{To explore an AIGC service provisioning framework in wireless networks, it is required to meet diverse service requirements under limited communication resources, unevenly distributed computational resources, and dynamic channel quality.
In this case, semantic communication (SemCom) is a potential technique to empower wireless AIGC provisioning.
SemCom aims to effectively transmit necessary semantic information of source message, as opposed to the accurate reception of every single binary bit without considering its meaning \cite{qin2021semantic, xia2023wiservr, liang2023vista, zhao2022background,sun2024s}.
In this way, the pressure of communication resources, especially bandwidth, can be relieved with a slim data size \cite{huang2021deep}. 
Meanwhile, the receiver with semantic decoder accurately interprets semantic information from distorted bits, thus improving transmission reliability at a semantic level and allowing for suitable AIGC services over harsh wireless channels \cite{guler2018semantic}. 
Furthermore, SemCom models and generative AI models usually exploit similar encoder and decoder structures and could involve overlapping neural network layers, which makes it natural to integrate these models without introducing additional system complexity \cite{liang2024generative}.}

However, resources in wireless networks, such as bandwidth, GPU/CPU cores, and computational power, are often with dynamic availability. 
Therefore, it is inefficient to directly integrate SemCom and AIGC with a fixed encoder and decoder structure, where the AIGC model is merely deployed within the encoder side or decoder side. 
This fixed structure makes the encoder/decoder unable to adjust the computing workloads according to the resource availability.
Furthermore, in the fixed structure of the encoder and decoder, the semantic density is unadjustable, causing a high semantic noise ratio under bad channel conditions, thereby posing additional denoising difficulty in decoding \cite{pan2024sc}.  
Meanwhile, users may have varying service requirements, such as latency and content quality, when accessing various AIGC services. 
Since the semantic computing workloads and semantic density are unadjustable in the transceiver, meeting these requirements becomes another difficulty.

In response to the unsolved difficulties, deploying diffusion models in both edge and local for the joint AIGC generation and transmission is a potential solution.
The diffusion models, as popular generative AI models, are utilized to denoise irrelevant information from pure Gaussian noise and finally retain valid information of content like text \cite{lin2023text}, audio \cite{huang2022prodiff}, or image \cite{rombach2022high}. 
By deploying diffusion models into the encoder and decoder, it is promising to develop a workload-adjustable transceiver, in which the semantic information can be generated within both edge and local. 
Additionally, machine learning (ML) is capable of intelligently deciding the denoising steps of the encoder and decoder as per resource availability and channel quality, i.e., optimizing the workload adaptation, thus meeting the content generation latency requirement. 
Furthermore, semantic density can also be optimized to reduce the semantic noise ratio, thus efficiently ensuring that the content quality is consistent and satisfies the requirements under varying channel conditions. 

\subsection{Contributions and Organization}
In this paper, to narrow down our focus, we delve specifically into image-based AIGC service provisioning over wireless networks. 
We first propose a SemCom-empowered AIGC (SemAIGC) framework, in which a workload-adjustable transceiver is deployed to jointly generate and transmit images, and a \uline{r}esource-aware w\uline{o}rkl\uline{o}ad \uline{t}rade-off (ROOT) scheme is developed to intelligently make workload adaptation decisions for transceiver.
Simulation results demonstrate the effectiveness and robustness of our proposed SemAIGC under dynamic resource availability, varying channel quality, and diverse service requirements.
The main contributions of the paper are summarized as follows:

\begin{itemize}
    \item To address communication and computational resource constraints, we exploit SemCom and AIGC to propose a wireless SemAIGC framework, which pre-trains models at the cloud, extracts and generates semantic information at the edge, and uses SemCom to transmit semantic information to local for fine-tuning and restructuring.
    \item We modify the diffusion models to design a workload-adjustable AIGC transceiver in the SemAIGC framework. In this way, the workload at both the transmitter and receiver sides can be adjusted by changing the denoising steps as per the available resources, channel condition, and service requirements.
    \item We theoretically prove that the modified diffusion model deployed at the receiver to denoise the semantic noise caused by the wireless channel. 
    \item In order to optimize the workload adaption in SemAIGC, we formulate the problem as a Markov decision process (MDP) by considering dynamic resource availability, dynamic channel quality, and diverse AIGC service requirements. We then propose a dueling double deep Q network (D3QN)-based ROOT scheme to solve the problem and intelligently make the workload adaptation decisions. 
\end{itemize}

This paper is organized as follows. 
In Section \ref{sec:Framework}, we present the SemAIGC framework, followed by the design of edge transmitter and local receiver in Section \ref{sec:Transceiver}.
The ROOT scheme design for the workload-adjustable transceiver is elaborated in Section \ref{sec:Schemne}.
Then, we evaluate the performance gain of the SemAIGC framework and ROOT scheme in Section \ref{sec:Simulation}.
Section \ref{sec:Conculsion} concludes the paper.

\begin{figure*}[htbp]
    \centering
    \includegraphics[width=0.98\textwidth]{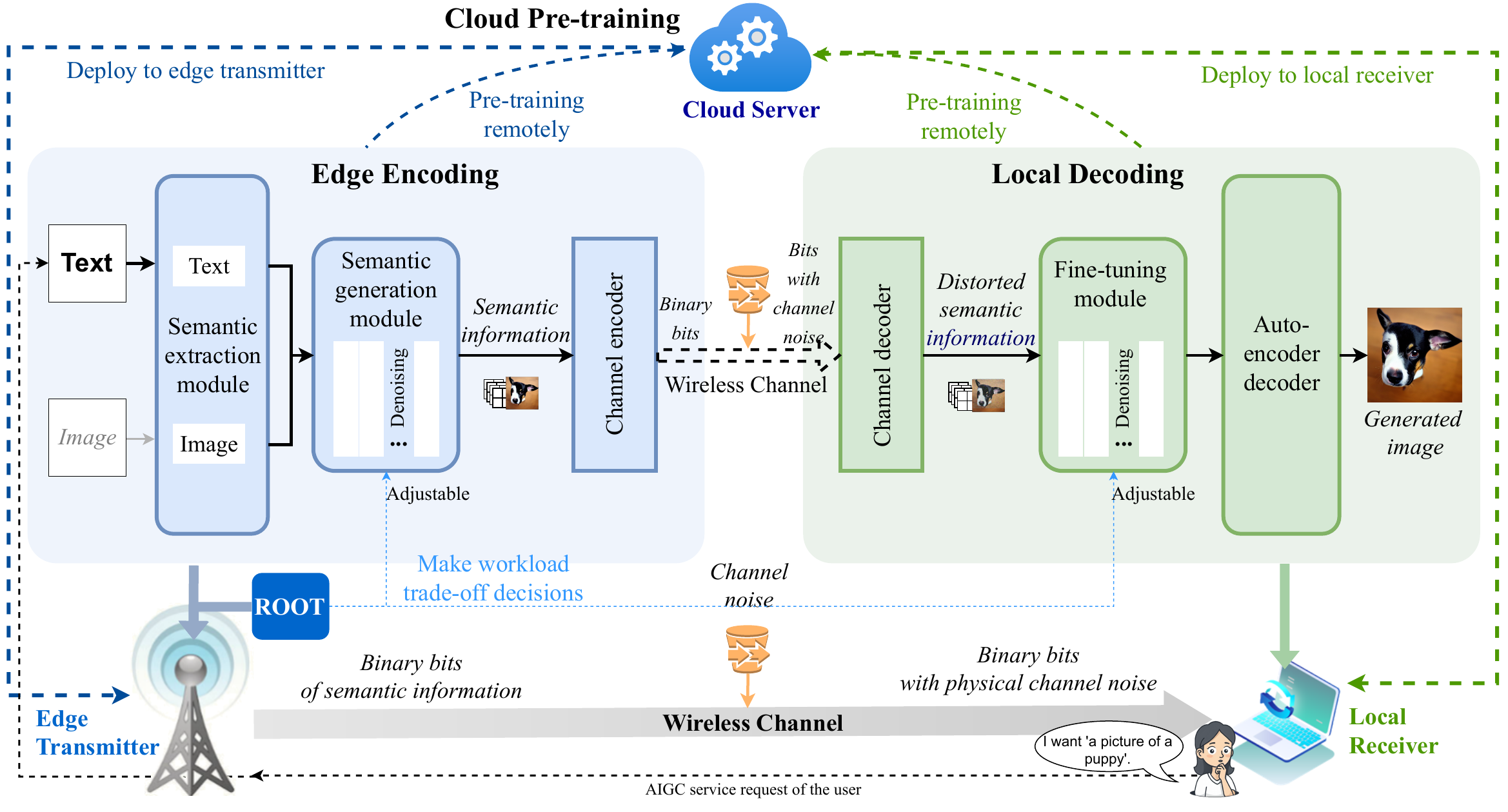}
    \caption{The proposed SemAIGC framework. \textbf{Stage I} is the cloud pre-training process that carried out in the cloud server to train transceiver semantic processing networks and ROOT scheme networks; \textbf{Stage II} edge encoding is executed by the edge transmitter, for example, a base station with an edge server, which is the major stage of semantic information generation and transmission; \textbf{Stage III} is the local decoding stage in the local receiver, deployed to fine-tune the semantic noise caused by channel noise and restructure semantic information into images.}
    \label{semanticmodel}
\end{figure*}

\section{SemCom-Empowered AIGC Generation and Transmission Framework}\label{sec:Framework}
Given unevenly distributed computational resources and limited communication resources, we introduce the SemAIGC framework, as shown in Fig. \ref{semanticmodel}. 
In this framework, our specific focus is on the generation and transmission of images from textual descriptions.
SemAIGC framework has three main stages, i.e., cloud pre-training, edge encoding, and local decoding. 

\subsection{Cloud Pre-Training} 
In SemAIGC, we utilize a cloud server to conduct the text-to-image (T2I) model training processes. 
Generally, a high-quality T2I model requires millions of parameters in a total size of 1-10 GB, stringent requirements are exposed for specialized hardware like GPU, CPU, and capable devices during its training. 
The cloud server as a centralized infrastructure with ample computing power, storage, database, etc. can be well-matched to the demands. 
As shown in Fig. \ref{semanticmodel}, the cloud server retains the copy of all the modules within both the encoder and decoder.
After feeding these modules with a substantial corpus of text-image pairs, the edge transmitter and local receiver can download the pre-trained encoder and decoder from the cloud server, respectively. 
\minew{Since the large models within the encoder-decoder generally perform stable after the initial offline pre-training, communication overhead could be very minor. 
In addition, even if periodical updates of the encoder and decoder are required, these models can be modularized as shown in Fig. \ref{semanticmodel}, and each update might merely cover partial modules and lead to less data transmission.}
Moreover, the neural network of the ROOT scheme can also be trained remotely and later deployed in the edge transmitter, then fine-tuned according to the local wireless network environment.
\subsection{Edge Encoding} 
As shown in Fig. \ref{semanticmodel}, the edge transmitter is composed of a semantic extraction module, a semantic generation module, and a channel encoder. 
The semantic extraction module comprises two independent networks, responsible for extracting semantic information from text and image, respectively.
\minew{As the core of the edge transmitter, the semantic generation module uses a diffusion model to predict and gradually denoise data starting from pure Gaussian noise according to the provided textual description. }
Compared with the pre-training stage, fewer computational resources are required in the AIGC generation and transmission stage. 
Therefore, the transmitter with edge computing servers is sufficient to run these modules with acceptable computing latency. 
Apart from the semantic encoder, the channel encoder in Fig. \ref{semanticmodel} is deployed at the edge transmitter to convert semantic information into binary bits for transmission.

\subsection{Local Decoding}
With the improvement of computing power in mobile devices, plenty of user devices are capable of executing information fine-tuning and recovery processes. 
In SemAIGC, the T2I decoder of a local receiver in Fig. \ref{semanticmodel} is composed of a semantic fine-tuning module, an autoencoder decoder module, and a channel decoder. 
Specifically, the channel decoder converts the binary bits into semantic information. 
\minew{Given the limited computing power of local devices, a lightweight diffusion model is deployed into the semantic fine-tuning module, which has similar functions as the aforementioned semantic generation module. }
After a rapid fine-tuning process, semantic noise can be eliminated, and then the autoencoder decoder module restructures semantic information into high-resolution images with text guidance.

\begin{figure*}[ht]
    \centering
    \includegraphics[width=0.8\textwidth]{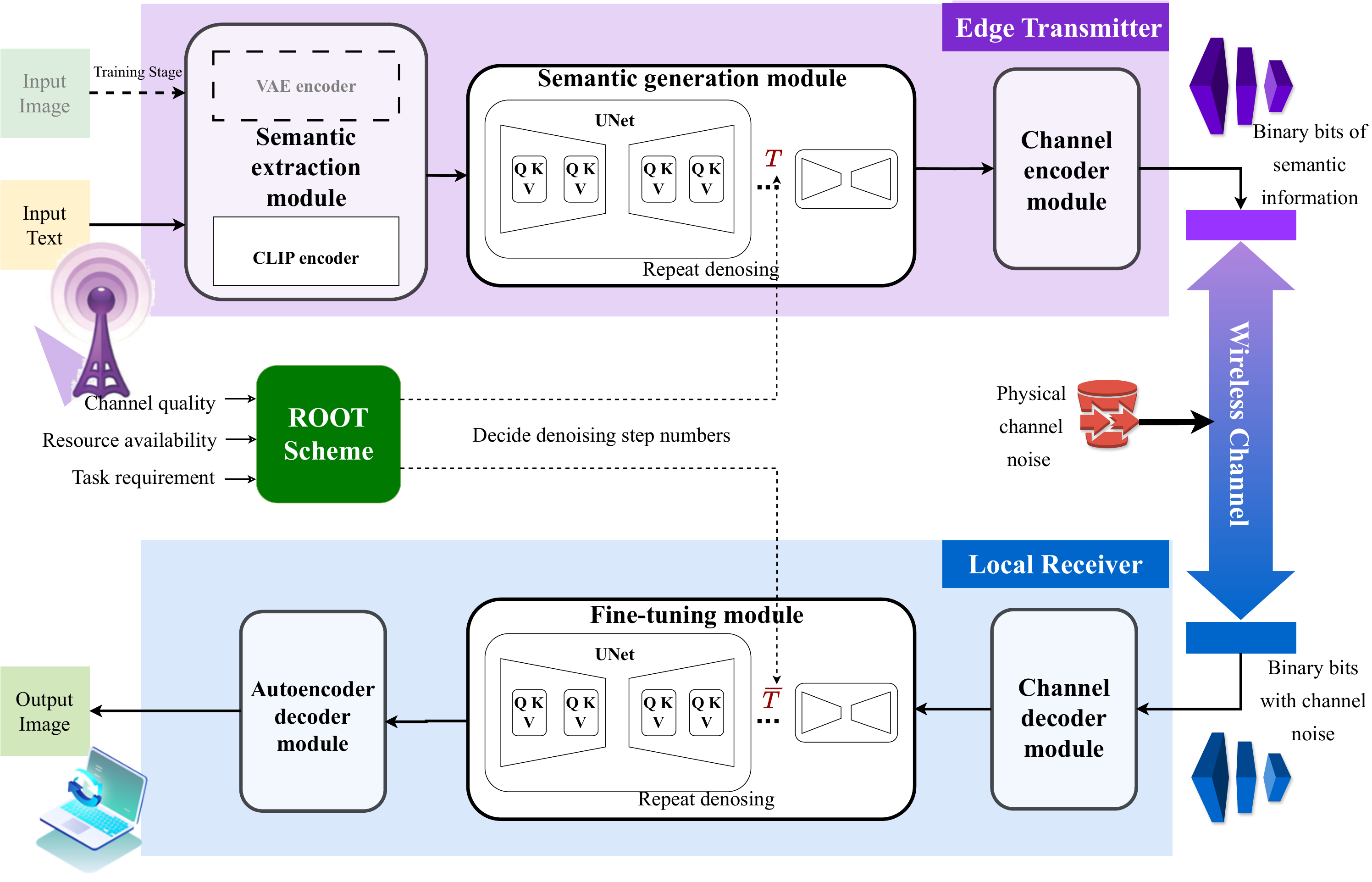}
    \caption{The workload-adjustable transceiver in SemAIGC. }
    \label{encoderdecoder}
\end{figure*}

\section{Workload-Adjustable Transceiver Design}\label{sec:Transceiver}
In this section, we consider a downlink transmission scenario, where the image semantic information is generated at the edge transmitter and then transmitted to the local receiver for further fine-tuning via a wireless channel. 
The edge transmitter and local receiver design in SemAIGC are detailed in the following.

\subsection{Encoder Design at Edge Transmitter}
As mentioned, the semantic extraction module, semantic generation module, and channel encoder are deployed in the edge transmitter, as shown in the upper part of Fig \ref{encoderdecoder}.

\textit{1) Semantic extraction module:} 
\minew{Let $\bm{s}_{\scriptscriptstyle t\!e\!x\!t} \in \mathbb{R}^{D}$ and $\bm{s} \in \mathbb{R}^{D_{H}\times D_{W}\times D_{C}}$ denote the input text and image to the semantic extraction module, respectively.
Here, $D$ is the number of tokens of input text $\mathbf{s}_{\scriptscriptstyle t\!e\!x\!t}$, and $D_{H}\times D_{W}\times D_{C}$ are the dimensions for a color image with $D_{C}$ the number of channels (e.g., 3 for the RGB image). 
Then, the semantic information extracted from the text and that from the image are denoted as $\mathbf{z}_{\scriptscriptstyle t\!e\!x\!t} \in \mathbb{R}^{d}$ and $\mathbf{z} \in \mathbb{R}^{d_{h}\times d_{w}\times d_{c}}$, respectively. 
Here, text semantic information is a vector of dimension $d$, and $d_{h}\times d_{w}\times d_{c}$ are the dimensions for image semantic information, typically $d_{h}\ll D_{H}, d_{w}\ll D_{W}$.
As the guidance of image generation, the semantic information of text  $\mathbf{z}_{\scriptscriptstyle t\!e\!x\!t}$ is extracted by a contrastive language-image pre-training (CLIP) encoder \cite{radford2021learning}.
Meanwhile, by using an attention-based variational autoencoder (VAE) model \cite{kingma2013auto}, the input image is encoded into the Gaussian-like distribution and outputs multiple sets of means and standard deviations. 
Then, we use these values to generate latent space image  $\mathbf{z}$, i.e., image semantic information. 
Note that the semantic information of real images is only used during the model training stage and is not directly involved in the T2I generation. 
The two modalities' semantic information is extracted as }
\begin{equation}
    \mathbf{z}_{\scriptscriptstyle t\!e\!x\!t}=\mathcal{E}\left(\mathbf{s}_{\scriptscriptstyle t\!e\!x\!t};\bm{\varphi }_{\scriptscriptstyle t\!e\!x\!t}\right),
    \mathbf{z}=\mathcal{E}\left(\mathbf{s};\bm{\varphi }\right).
\end{equation}   
Here, $\mathcal{E}\left(\cdot; \bm{\varphi }_{\scriptscriptstyle t\!e\!x\!t}\right)$ and $\mathcal{E}\left(\cdot; \bm{\varphi } \right)$ are the text encoder and image encoder with learnable parameters $\bm{\varphi }_{\scriptscriptstyle t\!e\!x\!t}$ and $\bm{\varphi }$, respectively. 

\textit{2) Semantic-generation module:} 

\begin{figure*}[ht]
    \centering
    \includegraphics[width=0.8\textwidth]{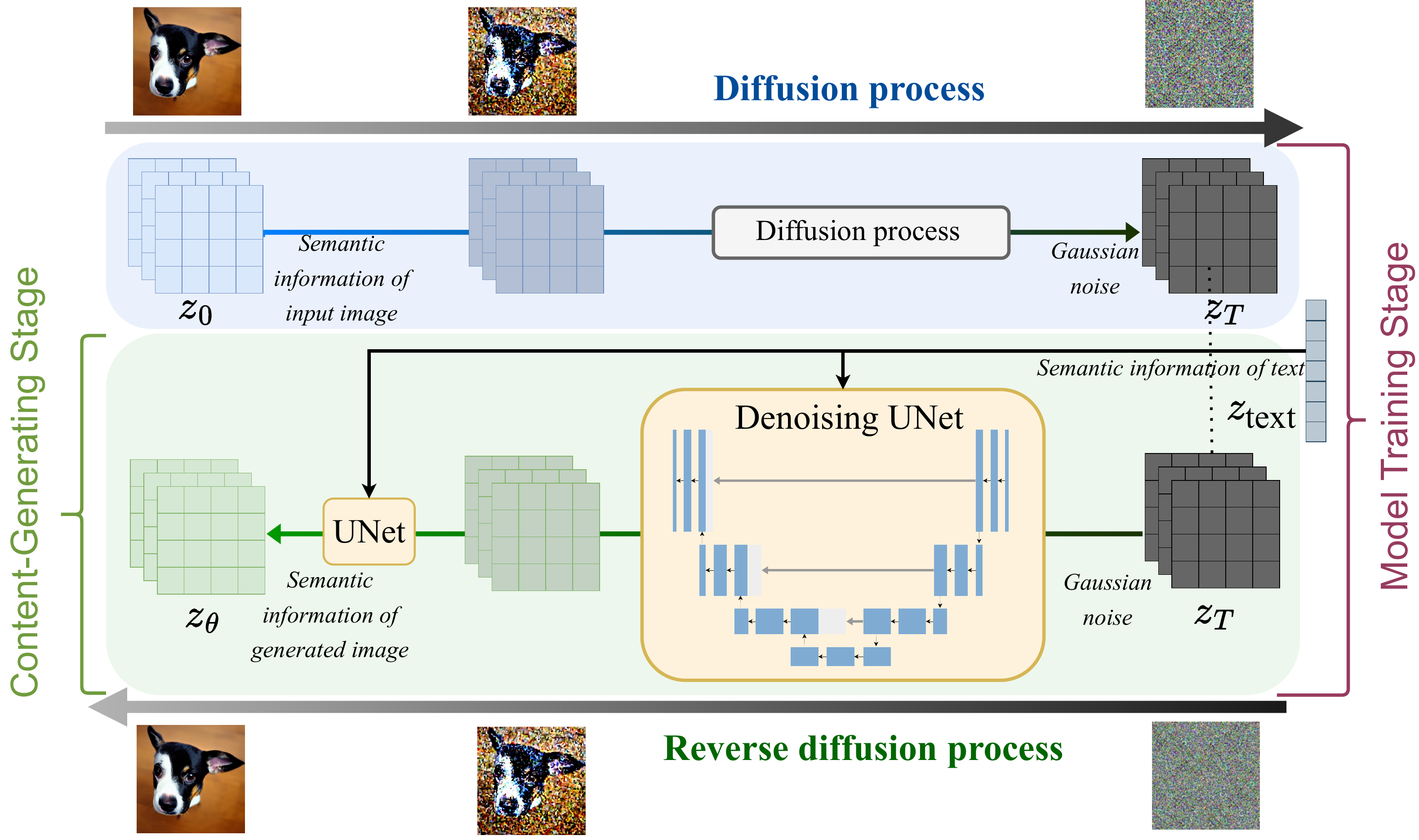}
    \caption{The diffusion and reverse diffusion processes in the semantic generation module. }
    \label{UNet}
\end{figure*}

The semantic generation modules in Fig. \ref{encoderdecoder} is based on the denoising diffusion probabilistic models (DDPMs), consisting of the diffusion process and the reverse diffusion process \cite{ho2020denoising}.
As shown in Fig. \ref{UNet}, the diffusion process is a fixed (or predefined) forward diffusion process, which is denoted as $q(\mathbf{z}_{t}|\mathbf{z}_{t-1})$. 
In the diffusion process, a scheduler gradually adds Gaussian noise at each time step $t \in [0,T]$, until the initial semantic information of an image $\mathbf{z}_{0}$ becomes pure noise $\mathbf{z}_{\scriptscriptstyle T}$, where $\mathbf{z}_{t} \in \mathbb{R}^{d_{h}\times d_{w}\times d_{c}}$. 
\minew{Therefore, the diffusion process is 
\begin{equation}
    \mathbf{z}_{t}=\sqrt{1-\beta_{t}}\mathbf{z}_{t-1}+\sqrt{\beta_{t}}\bm{\epsilon},
    \label{diffusion}
\end{equation}
where $0<\beta_{1}<\beta_{2}<\cdots<\beta_{\scriptscriptstyle T}<1$ is a known variance schedule with time-related constants, and $\bm{\epsilon}\sim  {\cal N}(0,\bm{I}), \bm{\epsilon}\in \mathbb{R}^{d_{h}\times d_{w}\times d_{c}}$ is the added Gaussian noise with an identity matrix $\bm{I}$.
}
Meanwhile, the reverse diffusion process in Fig. \ref{UNet} is a denoising process which is represented as $p(\mathbf{z}_{t-1}|\mathbf{z}_{t})$.  
In this process, a neural network, i.e., UNet \cite{ronneberger2015u}, is trained to learn the conditional probability distribution $p_{\bm{\theta}}(\mathbf{z}_{t-1}|\mathbf{z}_{t},\mathbf{z}_{\scriptscriptstyle t\!e\!x\!t})$ and gradually denoise the semantic information according to text semantic starting from pure noise until it ends up with the semantic information of an actual image. 
Here, $\bm{\theta}$ is the learnable parameters of the UNet, text guidance is implemented by concatenating the text embedding to the key-value pairs of each self-attention layer in the UNet \cite{rombach2022high}.
Accordingly, the reverse diffusion process is 
\begin{equation}
    \mathbf{z}_{t-1}=\frac{1}{\sqrt{\alpha_{t}}}\left(\mathbf{z}_{t}-\frac{1-\alpha_{t}}{\sqrt{1-\dot{\alpha_{t}}}}\bm{\epsilon}_{\bm{\theta}}\left(\mathbf{z}_{t},t,\mathbf{z}_{\scriptscriptstyle t\!e\!x\!t}\right)\right)+\bar{\sigma}_{t}\tilde{\bm{\epsilon}},
\end{equation}
where $\bm{\epsilon}_{\bm{\theta}}$ is the UNet predicted noise, $\alpha_{t}=1-\beta_{t}$ is a known constant of step $t$, $\dot{\alpha_{t}}=\prod_{i=1}^{t} \alpha_{i}$ is a cumulative product of $\alpha_{t}$, $\bar{\sigma}_{t}$ is a time dependent constant of step $t$, and $\tilde{\bm{\epsilon}} \sim {\cal N}(0,\bm{I})$ is the random normal Gaussian noise. 

Note that the semantic generation module conducts different processes during the pre-training stage and content-generating stage, as shown in Fig. \ref{UNet}.
\minew{In the pre-training stage, the semantic generation module executes both processes to learn the noise distribution, while only the reverse diffusion process is needed in content-generating stage.
In the content-generating stage, with pre-trained UNet, the target image is denoised from randomly generated pure noisy data $\mathbf{z}_{\scriptscriptstyle T}$. 
After the scheduled denoising steps $T$, the image semantic information is output according to text guidance, which is expressed as 
\begin{equation}
    \mathbf{z}_{\bm{\theta}}=\mathcal{F}_{1}(\mathbf{z}_{\scriptscriptstyle T}, T, \mathbf{z}_{\scriptscriptstyle t\!e\!x\!t};{\bm{\theta}}), 
\end{equation}
where $\mathcal{F}_{1}(\cdot;\bm{\theta})$ is the UNet with learnable parameter $\bm{\theta}$. }
The model training of diffusion models will be elaborated in Section \ref{subsec:Training}.

\textit{3) Channel encoder module:} 
\minew{As very few bits are required for text semantic information transmission, let us only focus on the transmission of image semantic information \cite{jiang2022deep}.
The transmitted signal is encoded as $\mathbf{Y}=\mathcal{C} ({\mathbf{z}_{\bm{\theta}}})$ via the channel encoder, where $\mathbf{Y} \in \mathbb{C}^{M_{c} \times k} $ is the analog signal, $k$ denotes the number of channels used to transmit the signal.

After passing through the physical channel, the received signal is expressed as
\begin{equation}
    \mathbf{Y}^{\prime}=\mathbf{H}\mathbf{Y}+\mathbf{N},
\end{equation}
where $\mathbf{H}\in\mathbb{C}^{M_{c}\times M_{c}} $ represents the channel gain matrix and $\mathbf{N}$ is the noise.
We consider the additive white Gaussian noise (AWGN) channel in this paper.\footnote{\minew{The used diffusion model is Gaussian noise-based hot diffusion,  which is designed to denoise Gaussian distributed noise. 
For filtering other types of noise, like Rayleigh noise,  it is necessary to specifically pre-train a cold diffusion-based module \cite{bansal2024cold}.}}
}
\subsection{Decoder Design at Local Receiver}
Then, we present the T2I decoder of the local receiver, which is composed of the channel decoder module, fine-tuning module, and autoencoder decoder module, as shown in the lower part of Fig. \ref{encoderdecoder}. 

\textit{1) Channel decoder module:}
The received binary bits are recovered as semantic information by a traditional channel decoder. 

The decoded information can be represented as 
\begin{equation}
    \begin{aligned}
        \mathbf{z}^{\prime}= \mathcal{C} ^{-1}(\mathbf{Y}^{\prime}), \\
    \end{aligned} 
    \label{decoder semantic}
\end{equation}
where $\mathcal{C}^{-1}$ is the channel decoder.
Therefore, $\bm{\epsilon}^{\cal C}=\mathbf{z}^{\prime}-\mathbf{z},  \bm{\epsilon}^{\cal C}\sim {\cal N}(0,\sigma^{2}\bm{I})$ is the \textit{semantic noise} with a variance matrix $\sigma^{2}\bm{I}$ caused by the physical channel noise $\mathbf{N}$.

\textit{2) Fine-tuning module:}
\minew{This is a lightweight version of the semantic generation module, which is used for fine-tuning decoded image semantic information. 
The reverse diffusion process of fine-tuning module is as follows
\begin{equation}
    \mathbf{z}_{\bar{\bm{\theta}}}=\mathcal{F}_{2}(\mathbf{z}^{\prime}, \bar{T}, \mathbf{z}_{\scriptscriptstyle t\!e\!x\!t};{\bar{\bm{\theta}}}), 
\end{equation}    
where $\mathcal{F}_{2}(\cdot;\bar{\bm{\theta}})$ is the UNet with learnable parameter $\bar{\bm{\theta}}$, $\bar{T}$ is the number of reverse diffusion steps for the fine-tuning module. 
Nevertheless, unlike the semantic generation module, the fine-tuning module considers semantic noise $\bm{\epsilon}^{\cal C}$ when training the UNet. }
We theoretically prove that semantic noise $\bm{\epsilon}^{\cal C}$ can be denoised according to the modified reverse diffusion process, as shown in the Proposition. \ref{Pro1}.

\begin{theorem}
    \minew{
    Assume \textit{semantic noise} $\bm{\epsilon}^{\cal C} \sim {\cal N}(0,\sigma^{2}\bm{I})$ is gradually added to semantic information, the reverse diffusion process from denoising step $t$ to $t-1$ in the fine-tuning module is  
    \begin{equation}
        \begin{aligned}
            &\mathbf{z}_{t-1}=\frac{1}{\sqrt{\alpha_{t}}}\left(\mathbf{z}_{t}-C(\alpha,\sigma,t)\bm{\epsilon}_{\bm{\theta}}(\mathbf{z}_{t},t,\mathbf{z}_{\scriptscriptstyle t\!e\!x\!t}) \right)+\bar{\sigma}_{t}\bm{\epsilon},\\
            where \; &C(\alpha,\sigma,t)=\frac{\left(1-\alpha_{t}+\sigma_{t,t-1}^{2} \right)\left(\sqrt{1-\dot{\alpha}_{t}}-\sigma_{t}^{2}\right)}{1-\dot{\alpha}_{t}+\sigma_{t-1}^{2}\alpha_{t}+\sigma_{t,t-1}^{2}}
        \end{aligned}
    \end{equation} 
    is a set of time related constants, and $\bm{\epsilon}_{\bm{\theta}}(\mathbf{z}_{t},t,\mathbf{z}_{\scriptscriptstyle t\!e\!x\!t}) $ is the noise predicted from semantic information $\mathbf{z}_{t}$. 
    Additionally, $\sigma_{t}$, $\sigma_{t-1}$, $\sigma_{t,t-1}$, and $\bar{\sigma}_{t}$  are time-dependent standard deviations for the random sample of Gaussian noise.
    }
    \label{Pro1}
\end{theorem}

\begin{IEEEproof}
    \minew{
    When the delivered semantic information is impacted by channel noise, according to (\ref{diffusion}), the semantic information is
    \begin{equation}
        \begin{aligned}
            \mathbf{z}_{t}&=\sqrt{\alpha_{t}}\mathbf{z}_{t-1}+\sqrt{1-\alpha_{t}}\bm{\epsilon}+\bm{\epsilon}^{\cal C}\\
            &=\sqrt{\dot{\alpha_{t}}}\mathbf{z}_{0}+\sqrt{1-\dot{\alpha_{t}}}\bm{\epsilon}+\bm{\epsilon}^{\cal C},\\ 
            &\bm{\epsilon}^{\cal C}\sim {\cal N} (0,\sigma^{2} \bm{I}), \bm{\epsilon}\sim {\cal N} (0,\bm{I}).
        \end{aligned}
    \end{equation}
    Here, semantic noise $\bm{\epsilon}^{\cal C}$ is added after transmission at one time, while in general diffusion, Gaussian noise is gradually added and denoised through multiple steps. 
    If denoising all semantic noise within a single step, minor noise prediction errors can lead to semantic information being lost or distorted. 
    Therefore, it is necessary to simulate the general diffusion process to gradually filter channel-caused semantic noise $\bm{\epsilon}^{\cal C}$. 
    We assume semantic noise is gradually added during transmission, thus vice it can be gradually removed.
    With diffusion steps $\{1,\cdots,t,\cdots,\bar{T}\}$, the diffusion process can be formulated as
    \begin{equation}
        \begin{aligned}
            \mathbf{z}_{t}&=\sqrt{\alpha_{t}}\mathbf{z}_{t-1}+\sqrt{1-\alpha_{t}}\bm{\epsilon}+\sqrt{\gamma_{t}}\bm{\epsilon}^{\cal C}\\
            &=\sqrt{\dot{\alpha}_{t}}\mathbf{z}_{0}+\sqrt{1-\dot{\alpha}_{t}}\bm{\epsilon}+\sigma \sum_{t=1}^{\bar{T}}\prod_{j=t+1}^{\bar{T}-1} \sqrt{\gamma_{t}} \sqrt{\alpha_{j}}\bm{\epsilon}, 
        \end{aligned}
        \label{ztz0}
    \end{equation}
    where $0<\gamma_{1}<\gamma_{2}<\cdots<\gamma_{t}<\cdots<\gamma_{\bar{T}}<1$ is a variance schedule with time related constants, and $\sum_{t=1}^{\bar{T}}\prod_{j=t+1}^{\bar{T}-1} \sqrt{\gamma_{t}} \sqrt{\alpha_{j}}=1$.
    
    The aim of fine-tuning module is to generate image semantic information $\mathbf{z}_{0}$ from received semantic information $\mathbf{z}_{t}$ via the reverse diffusion process.
    Therefore, it is important to predict the distribution probability of the semantic information in the previous step $\mathbf{z}_{t-1}$ with the present information $\mathbf{z}_{t}$, i.e., $q(\mathbf{z}_{t-1}|\mathbf{z}_{t},\mathbf{z}_{0})$.
    According to the Bayes' theorem, we have 
    \begin{equation}
        q(\mathbf{z}_{t-1}|\mathbf{z}_{t},\mathbf{z}_{0})=q(\mathbf{z}_{t}|\mathbf{z}_{t-1},\mathbf{z}_{0})\frac{q(\mathbf{z}_{t-1}|\mathbf{z}_{0})}{q(\mathbf{z}_{t}|\mathbf{z}_{0})}. \\   
        \label{equ11}
    \end{equation}
    The distributions are Gaussian, and according to equation (\ref{ztz0}), we have
    \begin{equation}
        \begin{aligned}
            &q(\mathbf{z}_{t}|\mathbf{z}_{0})\sim {\cal N} (\sqrt{\dot{\alpha}_{t}}\mathbf{z}_{0},1-\dot{\alpha}_{t}+{\sigma}_{t}^{2}),\\
            &q(\mathbf{z}_{t-1}|\mathbf{z}_{0})\sim {\cal N} (\sqrt{\dot{\alpha}_{t-1}}\mathbf{z}_{0},1-\dot{\alpha}_{t-1}+{\sigma}_{t-1}^{2}), \\
            &q(\mathbf{z}_{t}|\mathbf{z}_{t-1},\mathbf{z}_{0})\sim {\cal N} (\sqrt{\dot{\alpha}_{t}}\mathbf{z}_{t-1},1-\alpha_{t}+{\sigma}_{t,t-1}^{2}).
        \end{aligned}
    \end{equation}
    Here, noise variances ${\sigma}_{t}^{2}$, ${\sigma}_{t-1}^{2}$, and ${\sigma}_{t,t-1}^{2}$ can be determined according to $\sigma \prod_{j=t+1}^{\bar{T}-1} \sqrt{\gamma_{t}} \sqrt{\alpha_{j}}$.
    Then, with the Gaussian distribution formula $f(x)=\frac{1}{\sigma \sqrt{2\pi}}exp\left(-\frac{1}{2}\left(\frac{x-\mu}{\sigma}\right)^{2}\right)$, and as per equation (\ref{equ11}), we have
    \begin{equation}
        \begin{aligned}
            &q(\mathbf{z}_{t-1}|\mathbf{z}_{t},\mathbf{z}_{0})\varpropto\\ 
            &\frac{1}{\sigma \sqrt{2\pi}}exp\left(-\frac{1}{2}\left(\frac{\left(\mathbf{z}_{t}-\sqrt{\dot{\alpha}_{t}}\mathbf{z}_{t-1}\right)^{2}}{1-\alpha_{t}+{\sigma}_{t,t-1}^{2}} \right. \right.\\
            &\left.\left.+\frac{\left(\mathbf{z}_{t-1}-\sqrt{\dot{\alpha}_{t-1}}\mathbf{z}_{0}\right)^{2}}{1-\dot{\alpha}_{t-1}+{\sigma}_{t-1}^{2}}-\frac{\left(\mathbf{z}_{t}-\sqrt{\dot{\alpha}_{t}}\mathbf{z}_{0}\right)^{2}}{1-\dot{\alpha}_{t}+{\sigma}_{t}^{2}} \right)\right).  
        \end{aligned}
    \end{equation}
    Since the purpose of the reverse diffusion process is to predict the semantic information in the previous step, while $\mathbf{z}_{t}$ and $\mathbf{z}_{0}$ are known semantic information, we have
    \begin{equation}
        \begin{aligned}
            &q(\mathbf{z}_{t-1}|\mathbf{z}_{t},\mathbf{z}_{0})\varpropto\frac{1}{\sigma \sqrt{2\pi}}\times\\ 
            &exp\left(-\frac{1}{2}\left(\left(\frac{\alpha_{t}}{1-\alpha_{t}+{\sigma}_{t,t-1}^{2}}+\frac{1}{1-\dot{\alpha}_{t-1}+{\sigma}_{t-1}^{2}}\right)\mathbf{z}_{t-1}^{2}\right.\right.\\
            & \left.\left.-2\!\left(\frac{\sqrt{\alpha_{t}}\mathbf{z}_{t}}{1\!-\!\alpha_{t}\!+\!{\sigma}_{t,t-1}^{2}}\!+\!\frac{\sqrt{\dot{\alpha}_{t-1}}\mathbf{z}_{0}}{1\!-\!\dot{\alpha}_{t-1}\!+\!{\sigma}_{t-1}^{2}}\right)\!\mathbf{z}_{t-1}\!+\!C\left(\mathbf{z}_{t},\mathbf{z}_{0}\right)\!\right)\right).  
        \end{aligned}
    \end{equation}
    As Gaussian distribution formula can be transformed as $f(x)=\frac{1}{\sigma \sqrt{2\pi}}exp\left(-\frac{1}{2}\left(\frac{1}{\sigma^{2}}x^{2}-\frac{2\mu}{\sigma^{2}}x+\frac{\mu^{2}}{\sigma^{2}}\right)\right)$, 
    we have $\frac{\mu}{\sigma^{2}}=\left(\frac{\sqrt{\alpha_{t}}\mathbf{z}_{t}}{1\!-\!\alpha_{t}\!+\!{\sigma}_{t,t-1}^{2}}\!+\!\frac{\sqrt{\dot{\alpha}_{t-1}}\mathbf{z}_{0}}{1\!-\!\dot{\alpha}_{t-1}\!+\!{\sigma}_{t-1}^{2}}\right)$ and $\frac{1}{\sigma^{2}}=\left(\frac{\alpha_{t}}{1-\alpha_{t}+{\sigma}_{t,t-1}^{2}}+\frac{1}{1-\dot{\alpha}_{t-1}+{\sigma}_{t-1}^{2}}\right)$. 
    Then, we can calculate the mean of the distribution $q(\mathbf{z}_{t-1}|\mathbf{z}_{t},\mathbf{z}_{0})$ as
    \begin{equation}
        \begin{aligned}
        \bar{\mu}(\mathbf{z}_{t},\mathbf{z}_{0})\!=\frac{\sqrt{\alpha_{t}}\left(1-\dot{\alpha}_{t-1}+\sigma_{t-1}^{2}\right)}{1-\dot{\alpha}_{t}+\sigma_{t-1}^{2}\alpha_{t}+\sigma_{t,t-1}^{2}}\mathbf{z}_{t}+\\
        \frac{\sqrt{\dot{\alpha}_{t-1}}\left(1-\alpha_{t}+\sigma_{t,t-1}^{2}\right)}{1-\dot{\alpha}_{t}+\sigma_{t-1}^{2}\alpha_{t}+\sigma_{t,t-1}^{2}}\mathbf{z}_{0}.
        \end{aligned}
    \end{equation}
    Then, since (\ref{ztz0}) can be reversed as $\mathbf{z}_{0}=\frac{1}{\sqrt{\dot{\alpha}_{t}}}\left(\mathbf{z}_{t}-(\sqrt{1-\dot{\alpha}_{t}}-\sigma_{t}^{2})\bm{\epsilon}_{t} \right), \bm{\epsilon}_{t}\sim {\cal N}(0,\bm{I}) $, thus we have
    \begin{equation}
        \begin{aligned}
            \bar{\mu}(\mathbf{z}_{t},t)\!&= \frac{\sqrt{\alpha_{t}}\left(1-\dot{\alpha}_{t-1}+\sigma_{t-1}^{2}\right)}{1-\dot{\alpha}_{t}+\sigma_{t-1}^{2}\alpha_{t}+\sigma_{t,t-1}^{2}}\mathbf{z}_{t}+\\
            &\frac{1-\alpha_{t}+\sigma_{t,t-1}^{2}}{{\sqrt{\alpha}_{t}}\left(1-\dot{\alpha}_{t}+\sigma_{t-1}^{2}\alpha_{t}+\sigma_{t,t-1}^{2}\right)}\mathbf{z}_{t}-\\
            &\frac{\left(1-\alpha_{t}+\sigma_{t,t-1}^{2}\right)\left(\sqrt{1-\dot{\alpha}_{t}}-\sigma_{t}^{2}\right)}{{\sqrt{\alpha}_{t}}\left(1-\dot{\alpha}_{t}+\sigma_{t-1}^{2}\alpha_{t}+\sigma_{t,t-1}^{2}\right)}\bm{\epsilon}_{t}\\
            &=\!\frac{1}{\sqrt{\alpha_{t}}}\left(\mathbf{z}_{t}\!-\!\frac{\left(1\!-\!\alpha_{t}\!+\!\sigma_{t,t-1}^{2} \right)\left(\sqrt{1\!-\!\dot{\alpha}_{t}}\!-\!\sigma_{t}^{2}\right)}{1-\dot{\alpha}_{t}+\sigma_{t-1}^{2}\alpha_{t}+\sigma_{t,t-1}^{2}}\bm{\epsilon}_{t} \right). 
        \end{aligned}
   \end{equation} 
    Therefore, we have $\mu_{\bm{\theta}}(\mathbf{z}_{t},t)$ as the mean of distribution $q(\mathbf{z}_{t-1}|\mathbf{z}_{t},\mathbf{z}_{0})$. 
    Using a neural network to estimate $\bm{\epsilon}_{t}$, and according to the reparameterization trick, we have $\mathbf{z}_{t-1}=\frac{1}{\sqrt{\alpha_{t}}}\left(\mathbf{z}_{t}-C(\alpha,\sigma,t)\bm{\epsilon}_{\bm{\theta}}(x_{t},t) \right)+\bar{\sigma}_{t}\bm{\epsilon}$.
    }
\end{IEEEproof}

\textit{3) Autoencoder decoder module:}
This is the decoder of the VAE model with KL loss \cite{kingma2016improved}, which paints the final image output using the fused semantic information of text and image. 
The image semantic information is gradually recovered to the target image.
The final image from the autoencoder decoder module is output as
\begin{equation}
    \bar{s}=\mathcal{D} (\mathbf{z}_{\bar{\bm{\theta}}}|\mathbf{z}_{\scriptscriptstyle t\!e\!x\!t};\bar{\bm{\varphi}}),
\end{equation}
where $\mathcal{D} (\cdot;\bar{\bm{\varphi}})$ is the autoencoder decoder with learnable parameter $\bar{\bm{\varphi}}$.

\subsection{Training Algorithm}\label{subsec:Training}
\minew{
The training of image process modules is the key to generating a high-quality image from text. 
Specifically, the image semantic extraction module and the autoencoder decoder module are jointly trained according to the loss function of the VAE encoder-decoder \cite{kingma2013auto,kingma2016improved} with a dataset $\mathbf{S}=\{\mathbf{s}^{(1)},\cdots,\mathbf{s}^{(i)},\cdots\mathbf{s}^{(M)}\}$, $M$ is the dataset size. 
In the consideration of semantic similarity and reconstruction likelihood of each image $\mathbf{s}^{(i)}$, the loss function is 
\begin{equation}
    \begin{aligned}
        loss_{1}(\bm{\varphi} ,\bar{\bm{\varphi}})=&\frac{1}{2m}\!\sum_{i=1}^{M}\sum_{j=1}^{K}\!\left(1\!+\!\log \sigma_{i,j}^{2}\!-\!\sigma_{i,j}^{2}\!-\!\mu_{i,j}^{2}\right)\\
        &\!+\!\frac{1}{m}\sum_{i=1}^{M}\left(\|\mathbf{s}^{(i)}\!-\!\bar{\mathbf{s}}^{(i)}\|_{2}^{2}\right)\!,
    \end{aligned}   
    \label{imagesim}
\end{equation}
where $K$ is the set size of means $\mu_{i,j} \in \bm{\mu}_{i}$, and standard deviations $\sigma_{i,j}\in \bm{\sigma}_{i}$ generated for datapoint $\mathbf{s}^{(i)}$ by the VAE encoder. 
Here, the semantic similarity term $\sum_{j=1}^{K}\!\left(1\!+\!\log \sigma_{i,j}^{2}\!-\!\sigma_{i,j}^{2}\!-\!\mu_{i,j}^{2}\right)$ is the KL divergence between the approximate Gaussian postier $\mathcal{N}(\bm{\mu}_{i},\bm{\sigma}_{i}^2)$ and standard normal prior $\mathcal{N}(0,\bm{I})$.
Meanwhile, $\left(\|\mathbf{s}_{i}\!-\!\bar{\mathbf{s}}_{i}\|_{2}^{2}\right)$ is the restructuration likelihood term, which measures the difference between the original image $\bar{\mathbf{s}}_{i}$ and the reconstructed image $\bar{\mathbf{s}}_{i}$.
}

To achieve the efficient generation and transmission of high-quality content, the pre-training of the diffusion model in the semantic generation module is the crucial part. 
The corresponding objective of the information generator is to predict the noise distribution in the extracted semantic information as follows
\begin{equation} 
    loss_{2}=\mathbb{E}_{\mathcal{E}(\mathbf{s}),{\bm{\epsilon}  }\sim  \mathcal{N}(0,\bm{I}),t} \left[\|{\bm{\epsilon} }-{\bm{\epsilon}  }_{\bm{\theta}}\left(\mathbf{z},t|\mathbf{z}_{\scriptscriptstyle t\!e\!x\!t}\right) \|_{2}^{2}\right].
    \label{eqloss}
\end{equation}

For the fine-tuning module, the training objective is similar to (\ref{eqloss}), except its UNet is trained to learn the distribution of the semantic noise plus additional channel noise-caused semantic noise. 

\section{Intelligent Workload Trade-off Scheme Design}\label{sec:Schemne}
Considering the dynamic nature of resource availability and channel quality in wireless networks, as well as varying user preferences, it becomes essential to flexibly adapt the transmitter and receiver workload, thus catering to the service needs effectively.

\subsection{T2I Service Evaluation Metrics}
In our SemAIGC framework, image quality and latency are the two crucial performance metrics that affect service quality.
The image quality is primarily influenced by semantic similarity and reconstruction likelihood. 
We can fetch the image quality score by evaluating the loss between the SemAIGC image and the target image\footnote{\minew{In practical use, if there is no target image available, the aesthetics scores predictors or user feedback mechanisms can be used to evaluate the image quality in model training and image generation \cite{hosu2019effective}.}}, as defined in (\ref{imagesim}). 
Meanwhile, the latency of accessing T2I service by the local receiver $j$ from edge transmitter $i$ consists of transmission delay $L_{1}$, edge computing delay $L_{2}$, and local computing delay $L_{3}$, i.e., $L=L_{1}+L_{2}+L_{3}$.
\minew{Let $O_{i,j}$ denote the data size of semantic information transmitted by edge transmitter $i$ to local receiver $j$, then the transmission delay is 
\begin{equation}
    \begin{aligned}
        &L_{1}=\frac{O_{i,j}}{v_{i,j}},
    \end{aligned}
    \label{transmission_delay}
\end{equation}
where $v_{i,j}=B_{i,j}\cdot \log_{2}(1+\mathit{S\!N\!R}_{i,j})$ is the bit transmission rate of the link from transmitter $i$ to receiver $j$, $\mathit{S\!N\!R}_{i,j}$ is the signal-to-noise-ratio (SNR) of this link, and $B_{i,j}$ represents the bandwidth allocated to this link. 
}

For the computing delay,  according to Amdahl's law, the computing delay consists of the computing time of the serialized part and the computing time of the parallelizable part \cite{luo2021resource}. 
In this work, the computing speed of edge transmitter $i$ is majorly dependent on the GPU\footnote{\minew{It is assumed that the GPUs work within an ideal environment with suitable device temperature, enough graphics memory, and ample processing memory. In practice, real-time monitoring techniques can be integrated into the proposed ROOT for accounting these environmental variations.}}. 
\minew{Meanwhile, the majority of the computing workload is caused by the reverse diffusion process of the semantic generation module and fine-tuning module, we denote the delay resulting from other modules as a small constant $l$ \cite{croitoru2023diffusion}. 
Additionally, since the D3QN model within ROOT merely involves a small-scale network, and can pre-make trade-off decisions within a few milliseconds, the additional computational complexity and latency in AIGC services are minimum. }
Therefore, the edge computing delay is calculated as 
\begin{equation}
    \begin{aligned}
        L_{2}&=\frac{(1-\tau)W_{i}(T) O_{\mathbf{s}}}{\nu_{i}}+\frac{\tau W_{i}(T) O_{\mathbf{s}}}{\nu_{i}m_{i}}+l \\
        &=\frac{W_{i}(T) O_{\mathbf{s}}}{\nu_{i}}\left(1-\tau+\frac{\tau}{m_{i}}\right)+l,
    \end{aligned}
    \label{Computing_delay}
\end{equation} 
\minew{where $W_{i}(T)$ is the processing density (in GPU/CPU cycles/bits) of the transmitter UNet that corresponds to the total denoising step number $T$ of the semantic generation module, $O_{\mathbf{s}}$ is the input data size, and $ \tau \in [0,1]$ stands for the parallelizable fraction of AIGC task. }
Moreover, $\nu_{i}$ is the processing capability (i.e., the GPU/CPU frequency in cycles/s) of each core of the GPU, and $m_{i}$ denotes the GPU core number of edge transmitter $i$. 
The local computing delay $L_{3}$ is calculated in the same way. 

The Proposition \ref{Pp1} proves that to allow SemAIGC to outperform the edge AIGC framework in terms of total latency, the information compression rate should fit certain conditions. 
Here, the edge AIGC framework generates images by the T2I model deployed in the edge server and delivers images with traditional communication.
\minew{
\begin{theorem}
    Assuming identical channel quality and resource allocation under the SemAIGC framework and the edge AIGC framework, to make the SemAIGC framework provide lower service latency than the edge AIGC framework, i.e., $L<L_{\scriptstyle edge}$, the compression rate of the AIGC encoder should satisfy  
    \begin{equation}
        \begin{aligned}
            \frac{O_{i,j}}{O_{\mathbf{s}}}<\left|1\!-\!\frac{W_{j}(\bar{T})v_{i,j}\left[C_{1}\nu_{i}\!-\!C_{2}\nu_{j}\right]}{\nu_{i}\nu_{j}\!+\!W_{j}(\bar{T})v_{i,j}\left[C_{1}\nu_{i}\!-\!C_{2}\nu_{j}\right]}\right|,
        \end{aligned}
    \end{equation}
    where $C_{1}$ and $C_{2}$ are constants, and $0<C_{1}<1$, $0<C_{2}<1$. 
    \label{Pp1}
\end{theorem}

\begin{IEEEproof}
    When the SemAIGC framework provides lower latency than the edge AIGC framework $L<L_{\scriptstyle edge}$, we have 
    \begin{equation}
        \begin{aligned}
            \frac{O_{i,j}}{v_{i,j}}\!+\!\frac{W_{i}(T)O_{\mathbf{s}}}{\nu_{i}}(1-\tau+\frac{\tau}{m_{i}})\!+\!\frac{W_{j}(\bar{T})O_{i,j}}{\nu_{j}}(1-\tau+\frac{\tau}{m_{j}})<\\
            \frac{O_{\mathbf{s}}}{v_{i,j}}+\frac{W_{i}(T)O_{\mathbf{s}}+W_{j}(\bar{T})O_{i,j}}{\nu_{i}}(1-\tau+\frac{\tau}{m_{i}}).
        \end{aligned}
        \label{L1L2}
    \end{equation}
    Then, equation (\ref{L1L2}) is transformed as follows
    \begin{equation}
        \begin{aligned}
        O_{i,j}\!\cdot\!\left(\frac{1}{v_{i,j}}\!+\!\frac{W_{j}(\bar{T})}{\nu_{j}}(1\!-\!\tau\!+\!\frac{\tau}{m_{j}})\!-\!\frac{W_{j}(\bar{T})}{\nu_{i}}(1\!-\!\tau\!+\!\frac{\tau}{m_{i}}) \right)< \\
        O_{\mathbf{s}}\cdot\frac{1}{v_{i,j}}.
        \end{aligned}
    \end{equation}
    Given that $\frac{O_{i,j}}{O_{\mathbf{s}}}>0$, and $\tau$, $m_{i}$ and $m_{j}$ are constants, we prove Proposition \ref{Pp1}.
\end{IEEEproof}
}
\subsection{Computing Workload Adaptation Problem Formulation}
\minew{
Let $\hat{L}_{i}$ be the lowest latency that edge AIGC or local AIGC can achieve for generate image in time $i$, $\Psi$ denotes the quality factor of the generated image, if the quality fail to meet the requirement, $\Psi=0$; otherwise, $\Psi=1$.
The optimization problem can be formulated as 
\begin{subequations}
    \begin{align}
        &\max_{T_{i},\bar{T}_{i}}\sum_{i=1}^{M_{t}}\Psi_{i} (\hat{L}_{i} - L_{i}), \label{eq:p}\\
        s.t.\; &T_{i} \in \mathbb{Z}_{\geq 0}, \forall i, \label{eq:pC1}\\
        &    \bar{T}_{i} \in  \mathbb{Z}_{\geq 0}, \forall i, \label{eq:pC2}\\
        &    T_{i}+\bar{T}_{i} > 0, \forall i, \label{eq:pC3}\\
        &    \Psi \in \{0,1\}, \forall i,\label{eq:pC4}
    \end{align}
    \label{eqp}
\end{subequations}
where constraints (\ref{eq:pC1})-(\ref{eq:pC3}) ensure that the denoising steps of the transmitter and receiver are positive integer or zero, and the total denoising step larger than 0. 
Constraint \ref{eq:pC4} restricts the quality factor of the generated image.

To solve this problem, we transform it into an MDP problem, and the action, state, reward, and objective function of this problem are defined as follows. 
}
\textit{Action:} In this work, the edge transmitter acts as the agent to decide its encoding workload. 
We denote $a\in [0,\hat{T}]$ as the action, which is the reverse diffusion steps taken charge by the edge transmitter's encoder.
Here, $\hat{T}$ is the largest reverse diffusion step for image generation.
Meanwhile, the number of denoising steps in the fine-tuning module is set to $\hat{T}-a$. 

\textit{State:} The state of system is denoted as $x=\{\bar{W}_{i},\bar{W}_{j},\overline{\mathit{S\!N\!R}}_{i,j},\bar{\nu}_{i},\bar{\nu}_{j},\bar{B}_{i,j},\bar{L}\}$, where $\bar{W}_{i}$ and $\bar{W}_{j}$ represent the average processing density of edge transmitter $i$ and local receiver $j$, respectively.
Moreover, $\bar{\nu}_{i}$ and $\bar{\nu}_{j}$ are the average processing capability of transmitter $i$ and receiver $j$, $\overline{\mathit{S\!N\!R}}_{i,j}$ denotes the average SNR of the link from transmitter $i$ to receiver $j$, $\bar{B}_{i,j}$ is the average bandwidth allocated to this link, and $\bar{L}$ represents the average delay requirement of receiver $j$. 
A D3QN model is sufficient to analysis this state space.

\textit{Reward:} \minew{In order to provide a required image quality with target latency according to the specific requirements of the receiver, the reward function is set as
\begin{equation}
    R= \left\{
    \begin{array}{lcr}
     1,& & {L\leq\check{L}},\\
     \frac{\eta_{L} \check{L}-L}{\eta_{L}\check{L}_{j}-\check{L}},& & {\check{L}\leq L<\eta_{L}\check{L}}, \\
     0,& & otherwise. 
    \end{array} \right.
    \label{eqreward}
\end{equation}
Here, $\check{L}_{j}$ is the delay requirement of receiver $j$, $\eta_{L}>1$ is the latency tolerance factor. }

\begin{algorithm}[htbp]
    \caption{The training the process of ROOT.}
    \label{alg1}
    \begin{algorithmic}[1]
    \STATE Initialize the parameter of two networks $\omega$ and $\bar{\omega}$, the discount rate $\lambda$, the minibatch size $m_{\scriptscriptstyle  \mathcal{D}}$, and time $t=0$;
    \FOR{Episode $1$ to the maximum training episode}
    \STATE Decide an action $a$ according to estimated Q-value under state $x$;
    \STATE Start the SemAIGC image generation
    \STATE Execute transmitter-side semantic processing according to action $a$;
    \STATE Transmit image semantic information;
    \STATE Execute receiver-side semantic processing according to the action; 
    \STATE Observe reward $r$, and update state $x^{\prime}$;
    \STATE Store transaction $(x,a,r,x^{\prime})$ into replay buffer; 
    \STATE Experience replay: Sample random minibatch of transactions $(x^{i},a^{i},r^{i},x^{i+1})$ from the replay buffer;
        \FOR{$i=1 $ to $ m_{\scriptscriptstyle  \mathcal{D}}$}
            \STATE Computer the target value $y^{i}=r^{i}+\lambda {\cal Q}_{\omega}\left(x^{i+1},argmax\left({\cal Q}\left(x^{i+1},a^{i+1};\omega^{i}\right)\right)\right)$;
            \STATE Update the Q-network $\omega\leftarrow \omega-\eta \cdot \frac{1}{m_{\scriptscriptstyle  \mathcal{D}}}\sum_{i\in m_{\scriptscriptstyle  \mathcal{D}}}\left[y^{i}-{\cal Q}\left(x^{i},a^{i}\right)\right]\cdot \nabla_{\omega}{\cal Q}_{\omega}\left(x^{i},a^{i}\right)$.
        \ENDFOR    
    \ENDFOR
    \end{algorithmic}
\end{algorithm}

\subsection{ROOT Scheme}
To solve the formulated MDP problem, we propose the ROOT scheme to dynamically adjust image quality and service latency by intelligently making computing workload adaptation decisions. 
Given the discrete action space in this scenario, a suitable approach is to employ a value-based DRL method. 
The D3QN has demonstrated remarkable performance in solving problems with discrete action spaces \cite{wang2016dueling}. 
It effectively mitigates the issue of overestimation of action values in the Q-learning model and can be further enhanced by incorporating the decaying $\varepsilon$-greedy strategy in the policy learning. 
Therefore, we apply the D3QN in ROOT to estimate the optimal action value and efficiently make the optimal workload adaptation decisions. 

The D3QN is composed of two networks.
One network with parameter $\omega$ is used to estimate the state value function $V(x)$ and the other with parameter $\bar{\omega}$ is used to represent the state-dependent action advantage function $ \varLambda(x, a)$ \cite{wang2016dueling}, the outputs of the two separate networks are integrated as
\begin{equation}
    \begin{aligned}
        {\cal Q}(x,a)=&V(x;\omega)+ \varLambda (x,a;\bar{\omega})\\
        & -\frac{1}{|A|}\sum_{a^{\prime}} \varLambda (x,a^{\prime};\bar{\omega}),
    \end{aligned}
\end{equation} 
where $a^{\prime}$ is the next action.

For action selection, the $\varepsilon$-greedy strategy is utilized to balance the exploration and exploitation where the transmitter selects an action $a$ randomly with probability $\varepsilon$ or selects an action $a=argmax_{a} {\cal Q}(x,a;\omega)=argmax_{a}\varLambda (x,a;\bar{\omega})$ with probability $1-\varepsilon$. 
After intelligent workload adaptation and AIGC service delivery, the state, action, reward, and the next state $(x,a,r,x^{\prime})$ are stored as transactions in the replay buffer for model training.
In each training iteration, there are $m_{\scriptscriptstyle  \mathcal{D}}$ transactions used in training, and the target is computed as
\begin{equation}
    y^{i}=r^{i}\!+\!\lambda {\cal Q}_{\omega}\!\left(x^{i+1},argmax\left({\cal Q}\left(x^{i+1},a^{i+1};\omega^{i}\right)\right)\right)\!, i\in m_{\scriptscriptstyle  \mathcal{D}},
\end{equation}
where $\lambda$ is a discount rate and $m_{D}$ is the batch size.
The Q-network is updated according to a gradient descent step
\begin{equation}
    \omega\leftarrow \omega- \eta \cdot \frac{1}{m_{\scriptscriptstyle  \mathcal{D}}}\sum_{i=1}^{m_{\scriptscriptstyle  \mathcal{D}}}\left[y^{i}-{\cal Q}\left(x^{i},a^{i}\right)\right]\cdot \nabla_{\omega}{\cal Q}_{\omega}\left(x^{i},a^{i}\right), 
\end{equation} 
where $\eta $ is the learning rate.
The training process of the ROOT scheme is presented in Algorithm \ref{alg1}.

With the D3QN, the transmitter can intelligently decide the denoising step numbers of the semantic generation module, so as that of the fine-tuning module. 

\section{Simulation Results and Discussions}\label{sec:Simulation}
In this section, we conduct simulations to evaluate the performance of the proposed SemAIGC framework and its ROOT scheme under different scenarios.
In order to demonstrate the rationality of integrating SemCom into AIGC, we compare with the following four benchmarks.
\begin{enumerate}
    \item Non-ROOT SemAIGC: This framework is identical to SemAIGC, except that the ROOT scheme is disabled.
    \item Non-fine-tuning AIGC: This framework offers a straightforward application of SemCom to the AIGC transmitter-receiver without a fine-tuning module in the receiver. 
    \item Edge AIGC: The image is generated by the T2I model deployed in edge server and uses traditional communication to deliver to users. 
    \item Local AIGC: This framework uses local receiver to compute all the processes of T2I generation, in which no wireless communication is involved. 
\end{enumerate}

\begin{figure*}[htbp]
    \centering
    \includegraphics[width=0.75\textwidth]{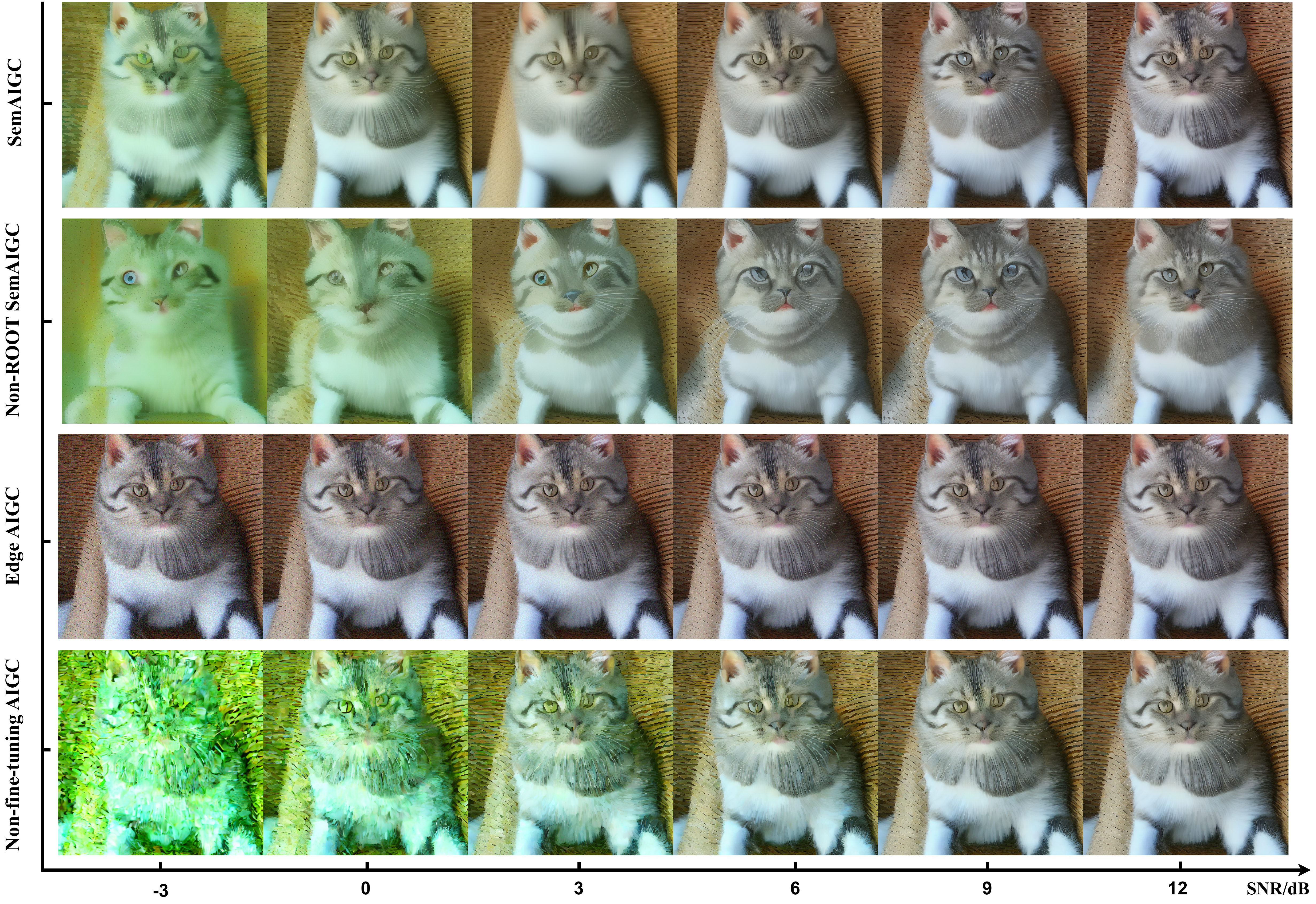}
    \caption{The delivered images of three AIGC generation and transmission frameworks under different SNRs (text input: ``A cute furry cat", the seed value for initial noise: 30).}
    \label{fig.Sim1}
\end{figure*}

\begin{figure}[htbp]
    \centering
    \includegraphics[width=0.35\textwidth]{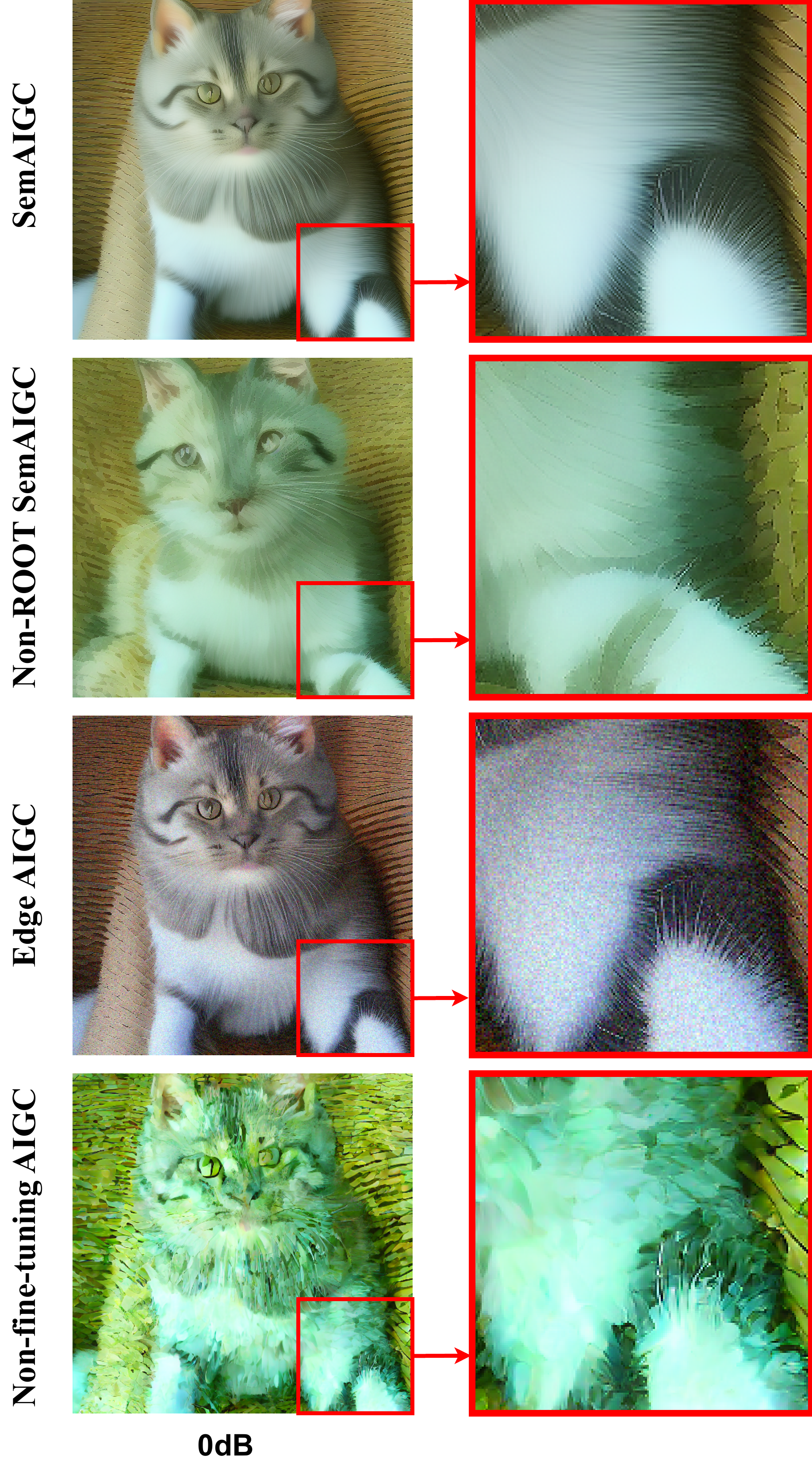}
    \caption{The image details of three AIGC generation and transmission frameworks under 0dB wireless channel.}
    \label{fig.Sim2}
\end{figure}

\subsection{Simulation Settings}
\minew{In the simulation, a workstation with the GPU RTX A6000 acts as the edge transmitter and a laptop with the GPU of GTX1080 acts as the local receiver.
The bandwidth budget of the link from the transmitter to the receiver is $20 MHz$, and the SNR is set to a range of $[-6,15] dB$. 
The percentage of available computing resources (PACR) is adjusted by controlling the GPU processing time slot. 
Additionally, the required latency of the AIGC service is randomly generated within the range of $[5,25] s$, while the maximum latency is twice the minimum latency. 
The satisfaction score is the average reward in (\ref{eqreward}) for generating multiple images, in the range $[0,1]$.
If the generation time per image exceeds the default value of $60 s$, the task is failed. 

The model training is composed of two major parts, i.e., SemAIGC model training and ROOT scheme training. 
For the SemAIGC model, the \textit{Stable Diffusion v1-5}  model checkpoint is used in the pre-trained model \cite{rombach2022high}. 
The corresponding semantic extraction module of text is deployed with pre-trained OpenAI CLIP \cite{radford2021learning}.
Meanwhile, the pre-trained VAE encoder and decoder are used as the image semantic extraction module and autoencoder decoder module, respectively \cite{kingma2016improved}.
Moreover, for evaluating the quality of generated image quality, the \textit{laion-AI/aesthetic-predictor} is used to evaluate the aesthetic score.
To adjust the UNet weights of the fine-tuning module \cite{hu2021lora}, a subset within the \textit{laion-aesthetics v2 5+} is used, which contains high-resolution images, and corresponding caption texts. 
In training, all the training images are filtered to a size of $512\times 512\times 3$. 
All the pre-processed images are extracted as $1\times4\times64\times 64$ image semantic information by VAE encoder for UNet training. 
Moreover, the UNet is trained with 1 middle block in $8\times 8$ resolution, 12 encoding blocks, and 12 decoding blocks in 4 resolutions, i.e., $64\times 64$, $32\times 32$, $16\times 16$, and $8\times 8$ \cite{rombach2022high}.

For the D3QN setting in the ROOT scheme, both the DQN network and target network have an input layer with 7 neurons and a 20-neural output layer, as well as 2 hidden layers with each of 256 neurons \cite{wang2016dueling}.
We employ $ReLU$ as the activation function between the input layer and the two hidden layers.
The training minibatch size is set to 100, and the learning rate is set to 0.0003. 
For controlling action exploration, the exploration probability is initially set to $1$, and decays $1*10^{-4}$ each episode, while the minimum is $0.01$.
}

\begin{figure}[htbp]
    \centering
    \includegraphics[width=0.4\textwidth]{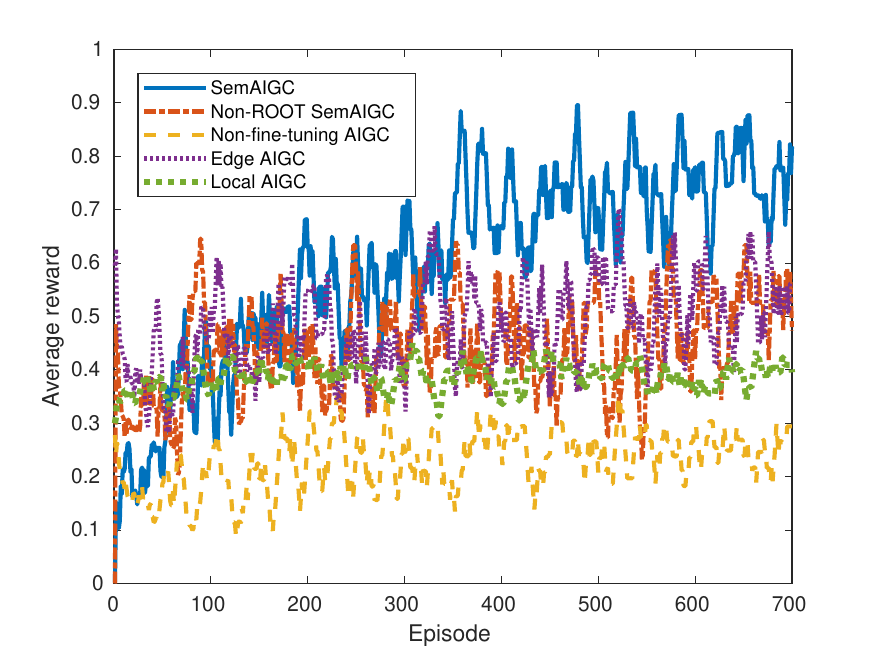}
    \caption{Convergence performance for four AIGC generation and transmission frameworks.}
    \label{fig.Sim3}
\end{figure}

\begin{figure}[htbp]
    \centering
    \begin{subfigure}[h]{0.4\textwidth}
        \centering
        \includegraphics[width=\textwidth]{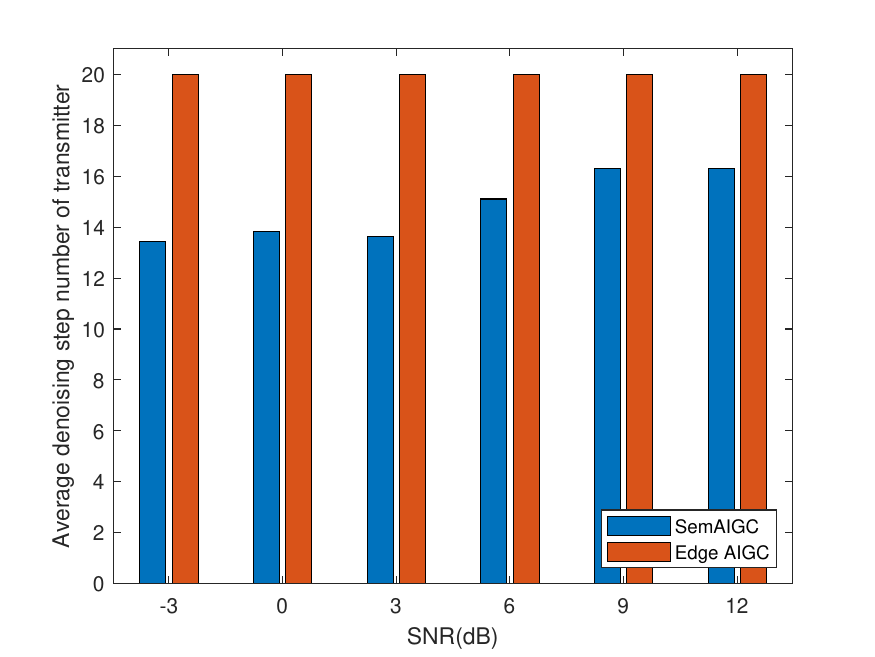}
        \caption{Average denoising steps executed by transmitter under varying SNRs.}
        \label{figadd2.1}
    \end{subfigure}
    \begin{subfigure}[h]{0.4\textwidth}
        \centering
        \includegraphics[width=\textwidth]{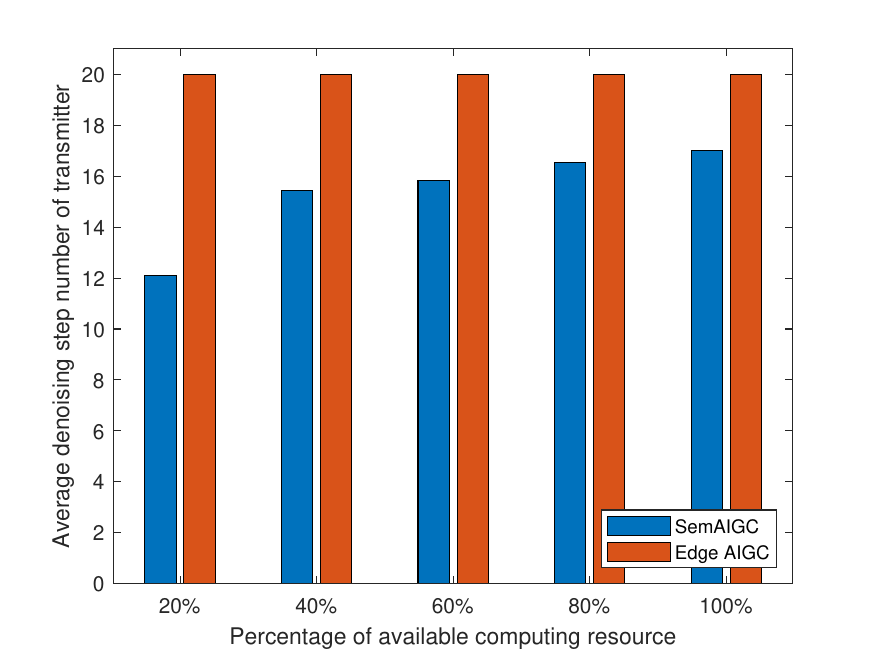}
        \caption{Average denoising steps executed by transmitter under different percentages of computing power.}
        \label{figadd2.2}
    \end{subfigure}
    \caption{Average denoising steps executed by transmitter.}
    \label{figadd2}
\end{figure}

\subsection{Numerical Results}

\begin{figure}[htbp]
    \centering
    \includegraphics[width=0.43\textwidth]{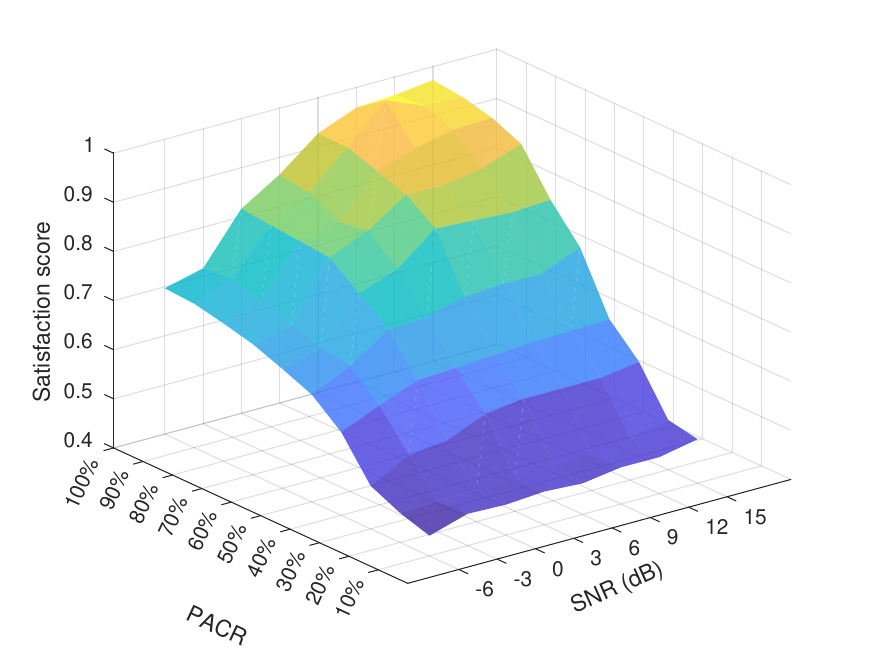}
    \caption{Satisfaction score for the SemAIGC framework under dynamic computing power and channel quality.}
    \label{fig.Sim4}
\end{figure}

\begin{figure*}[h]
    \centering
    \begin{subfigure}[h]{0.45\textwidth}
        \centering
        \includegraphics[width=\textwidth]{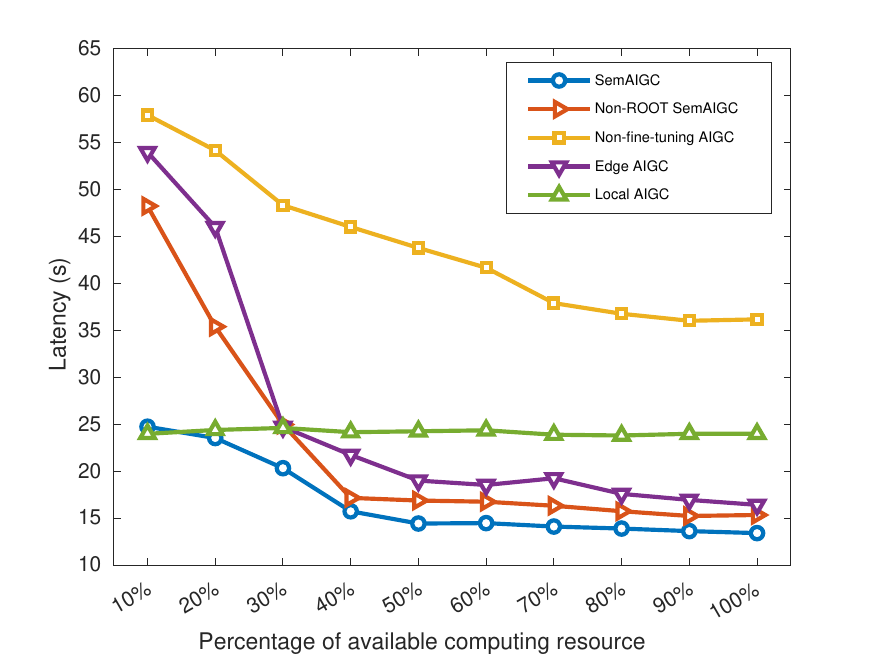}
        \caption{Average latency under different percentages of computing power.}
        \label{fig3.1}
    \end{subfigure}
    \begin{subfigure}[h]{0.45\textwidth}
        \centering
        \includegraphics[width=\textwidth]{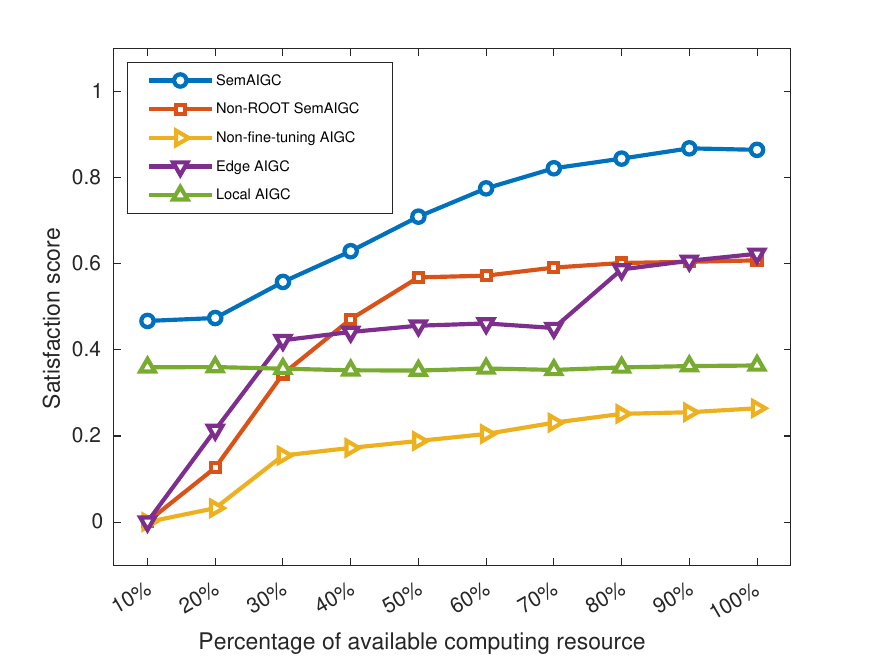}
        \caption{Satisfaction score under different percentages of computing power.}
        \label{fig3.2}
    \end{subfigure}
    
    \begin{subfigure}[h]{0.45\textwidth}
        \centering
        \includegraphics[width=\textwidth]{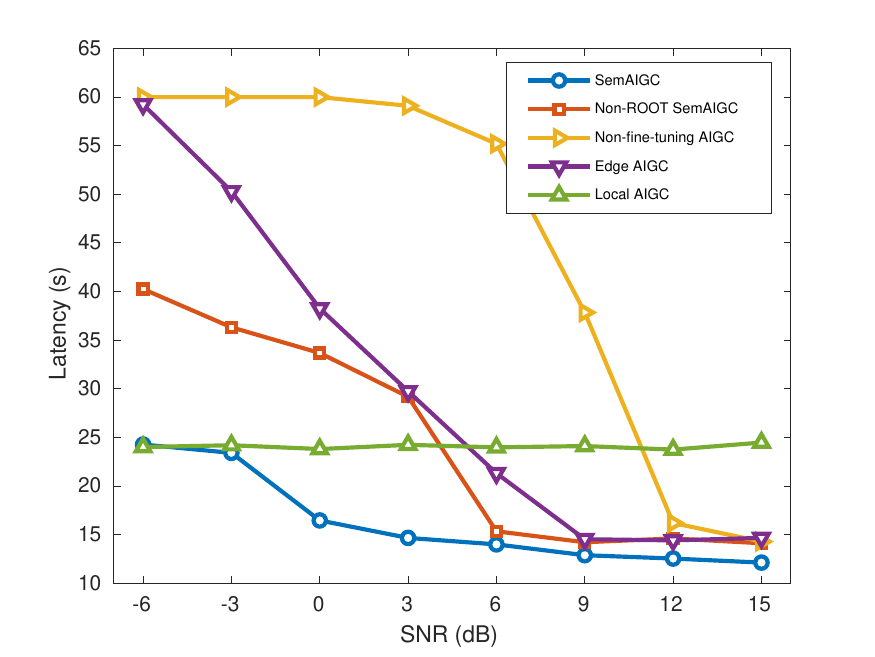}
        \caption{Average latency under different SNRs.}
        \label{fig3.3}
    \end{subfigure}
    \begin{subfigure}[h]{0.45\textwidth}
        \centering
        \includegraphics[width=\textwidth]{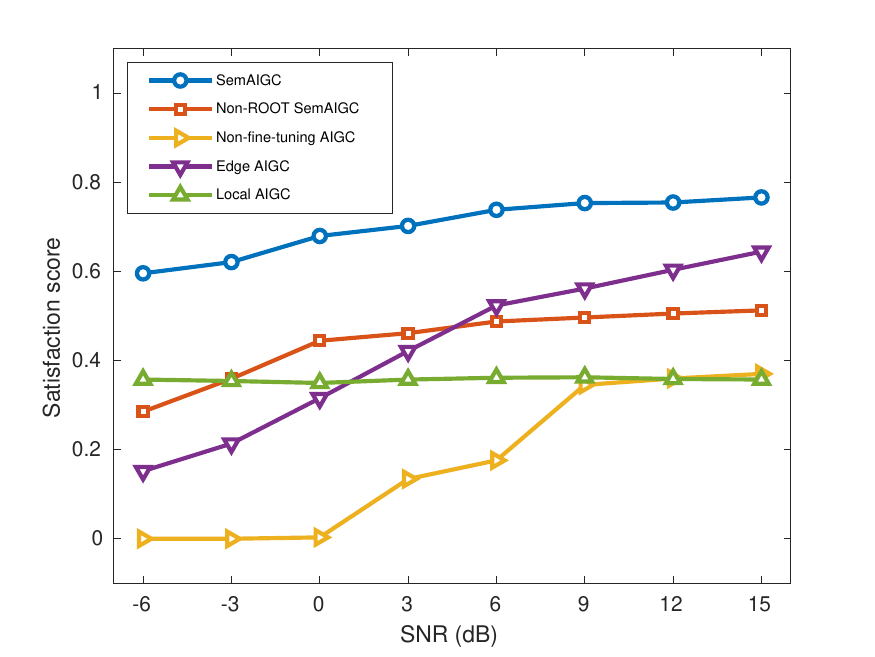}
        \caption{Satisfaction score under different SNRs.}
        \label{fig3.4}
    \end{subfigure}
    \caption{Service quality experienced by users for four different frameworks.}
    \label{fig.3}
\end{figure*}

We first examine the generated image quality under different channel qualities, shown in Fig. \ref{fig.Sim1} and Fig. \ref{fig.Sim2}. 
Notably, since Local AIGC does not involve wireless transmission, it is not considered as a benchmark here.
Moreover, as the number of denoising steps affects image quality, for fairness, we ought to set the total denoising step of different frameworks into the same number of 20.
The SemAIGC tends to intelligently adjust the denoising step of the transmitter and receiver.
Meanwhile, for the Non-ROOT SemAIGC service, both the denoising step of the semantic generation module and that of the fine-tuning module are set to 10, whereas the other two frameworks have 20 denoising steps solely in edge transmitter or local receiver. 
As shown in Fig. \ref{fig.Sim1}, it is obvious that the images generated by SemAIGC exhibit excellent clarity and align seamlessly with the textual description. 
The SemAIGC outperforms the Non-fine-tuning AIGC and Non-ROOT SemAIGC systems under all channel qualities while generating images with structures that nearly resemble those of the edge AIGC, thanks to the fine-tuning model's effective channel noise reduction and preservation of valuable semantic information. 
Since this study does not prioritize UNet training and its corresponding scheduler design, some impact from harsh channels may persist, resulting in a slight green tint in the images generated by SemAIGC. 
Nevertheless, this impact is significantly eliminated compared to images generated by Non-ROOT SemAIGC, because the majority of channel-caused semantic noise is eliminated by intelligently decided denoising step in SemAIGC.
Moreover, we compared the image details of SemAIGC and the other three frameworks under a $0$dB channel in Fig. \ref{fig.Sim2}. 
It is observed that SemAIGC can output the highest quality image with a smooth texture, while images from other frameworks have noisy points, blurred edges, and texture distortion. 
The results further indicate that the proposed SemAIGC is effective and robust in generating high-quality images over a noisy channel.  

Then, we evaluate its convergence performance by comparing its average reward with that of four alternative benchmarks, as shown in Fig. \ref{fig.Sim3}. 
It is worth noting that these four benchmarks employ the same reward function as the SemAIGC framework. 
\minew{From Fig. \ref{fig.Sim3}, we observe that although SemAIGC experiences moderate oscillations due to some randomly generated harsh service requirements, it still converges faster than other schemes, reaching convergence within approximately 350 episodes. }
This superior performance can be attributed to the ROOT scheme of SemAIGC, which intelligently adjusts the workload for the transmitter and receiver based on resource availability and channel quality. 
Fig. \ref{figadd2} demonstrates the effectiveness of the ROOT scheme for SemAIGC in terms of intelligently adjusting denoising steps, so as computing workload.
From Fig. \ref{figadd2.1} and Fig. \ref{figadd2.2}, we observe that the ROOT scheme of SemAIGC intelligently allocates more workload to the transmitter under batter channel conditions and higher computing resource availability, respectively. 

Additionally, Fig. \ref{fig.Sim4} illustrates the average satisfaction score of the user when utilizing the SemAIGC framework, in response to dynamic changes in channel quality and the PACR for the transmitter. 
Notably, the satisfaction score exhibits an increase, ranging from approximately $45\%$ to $98\%$, in tandem with the PACR and SNR.
Although under the most challenging conditions, characterized by the poorest channel quality with an SNR of $-6 dB$, SemAIGC can still attain an acceptable satisfaction score of approximately $70\%$, provided that the transmitter possesses sufficient computational resources. 
This impressive performance can be attributed to that the transmitter with ample computing power can speedily generate semantic information, and the fine-tuning module can effectively and promptly mitigate channel noise while preserving valuable semantic information.

To further demonstrate the performance of SemAIGC, we conduct a comparative analysis of its average latency and satisfaction score with other four frameworks, as depicted in Fig. \ref{fig.3}.
As seen in Fig. \ref{fig3.1}, when computational resources are fixed and the SNR varies dynamically, SemAIGC consistently maintains an average latency below $25 s$, surpassing the performance of the other three frameworks. 
This superiority is even more pronounced in Fig. \ref{fig3.2}. 
Although SemAIGC exhibits a slightly lower average latency than Edge AIGC by a mere $2 s$, its satisfaction score outperforms Edge AIGC by approximately $40\%$. 
This is due to the fact that SemAIGC prioritizes delivering images within a required time range rather than blindly pursuing extremely low latency.
Furthermore, Fig. \ref{fig3.3} and Fig. \ref{fig3.4} offer insights into the performance of these frameworks under varying SNRs. 
In Fig. \ref{fig3.3}, it becomes evident that under the harsh channel, the user with both the non-fine-tuning AIGC and Edge AIGC framework can experience significantly higher latency, approximately $25 s$ and $45 s$ greater than that of SemAIGC, respectively. 
This disparity arises from the absence of UNet, which is responsible for denoising received information from the receiver side and ensuring suitable image quality within the given time constraints.
Additionally, from Fig. \ref{fig3.4}, it is more obvious that the other four frameworks cannot provide a satisfying service as the SemAIGC, which further validates the effectiveness and robustness of SemAIGC.

\section{Conclusion}\label{sec:Conculsion}
In this paper, we introduce the SemAIGC framework as a novel approach for jointly generating and transmitting content in wireless networks. 
Our framework employs SemCom to reduce the consumption of communication resources. In addition, diffusion models are deployed into the encoder and decoder to develop a workload-adjustable transceiver and improve the utilization of computational resources.
Furthermore, a ROOT scheme is presented in SemAIGC to intelligently make workload adaptation decisions for providing adaptive AIGC access service. 
Simulation results demonstrate the superiority of SemAIGC framework in terms of service quality and robustness. 

This paper is a pioneer work in the realm of wireless AIGC provisioning. 
The utilization of SemCom shows instrumental in efficiently managing communication and computational resources in resource-intensive AIGC networks. 
Moreover, the generative AI model employed in this work can inspire the development of novel methods for semantic extraction, transmission, and recovery in the domain of image SemCom.

\section*{Acknowledgements}
This research of Dusit Niyato is supported by the National Research Foundation, Singapore, and Infocomm Media Development Authority under its Future Communications Research \& Development Programme, Defence Science Organisation (DSO) National Laboratories under the AI Singapore Programme (FCP-NTU-RG-2022-010 and FCP-ASTAR-TG-2022-003), Singapore Ministry of Education (MOE) Tier 1 (RG87/22), and the NTU Centre for Computational Technologies in Finance (NTU-CCTF). 
The work of Lan Zhang was partially supported by the US National Science Foundation CNS-2418308, CCF-2427316, and CCF-2426318. 

\normalem
\bibliographystyle{IEEEtran}
\bibliography{IEEEabrv,bibSemantic.bib}

\begin{thebibliography}{10}
\providecommand{\url}[1]{#1}
\csname url@samestyle\endcsname
\providecommand{\newblock}{\relax}
\providecommand{\bibinfo}[2]{#2}
\providecommand{\BIBentrySTDinterwordspacing}{\spaceskip=0pt\relax}
\providecommand{\BIBentryALTinterwordstretchfactor}{4}
\providecommand{\BIBentryALTinterwordspacing}{\spaceskip=\fontdimen2\font plus
\BIBentryALTinterwordstretchfactor\fontdimen3\font minus \fontdimen4\font\relax}
\providecommand{\BIBforeignlanguage}[2]{{%
\expandafter\ifx\csname l@#1\endcsname\relax
\typeout{** WARNING: IEEEtran.bst: No hyphenation pattern has been}%
\typeout{** loaded for the language `#1'. Using the pattern for}%
\typeout{** the default language instead.}%
\else
\language=\csname l@#1\endcsname
\fi
#2}}
\providecommand{\BIBdecl}{\relax}
\BIBdecl

\bibitem{jo2023promise}
A.~Jo, ``{The Promise and Peril of Generative AI},'' \emph{Nature}, vol. 614, no.~1, pp. 214--216, 2023.

\bibitem{wang2023survey}
Y.~Wang, Y.~Pan, M.~Yan, Z.~Su, and T.~H. Luan, ``{A Survey on ChatGPT: AI-Generated Contents, Challenges, and Solutions},'' \emph{arXiv preprint arXiv:2305.18339}, 2023.

\bibitem{roumeliotis2023chatgpt}
K.~I. Roumeliotis and N.~D. Tselikas, ``Chatgpt and open-ai models: A preliminary review,'' \emph{Future Internet}, vol.~15, no.~6, p. 192, 2023.

\bibitem{penalvo2022mobile}
F.~J.~G. Pe{\~n}alvo, A.~Sharma, A.~Chhabra, S.~K. Singh, S.~Kumar, V.~Arya, and A.~Gaurav, ``{Mobile Cloud Computing and Sustainable Development: Opportunities, Challenges, and Future Directions},'' \emph{International Journal of Cloud Applications and Computing (IJCAC)}, vol.~12, no.~1, pp. 1--20, 2022.

\bibitem{xu2023unleashing}
M.~Xu, H.~Du, D.~Niyato, J.~Kang, Z.~Xiong, S.~Mao, Z.~Han, A.~Jamalipour, D.~I. Kim, V.~Leung \emph{et~al.}, ``{Unleashing the Power of Edge-Cloud Generative AI in Mobile Networks: A Survey of AIGC Services},'' \emph{arXiv preprint arXiv:2303.16129}, 2023.

\bibitem{du2023enabling}
H.~Du, Z.~Li, D.~Niyato, J.~Kang, Z.~Xiong, D.~I. Kim \emph{et~al.}, ``{Enabling AI-Generated Content (AIGC) Services in Wireless Edge Networks},'' \emph{arXiv preprint arXiv:2301.03220}, 2023.

\bibitem{wang2023unified}
J.~Wang, H.~Du, D.~Niyato, J.~Kang, Z.~Xiong, D.~Rajan, S.~Mao \emph{et~al.}, ``{A Unified Framework for Guiding Generative AI with Wireless Perception in Resource Constrained Mobile Edge Networks},'' \emph{arXiv preprint arXiv:2309.01426}, 2023.

\bibitem{park2022real}
H.~Park, A.~Yessenbayev, T.~Singhal, N.~K. Adhikari, Y.~Zhang, S.~M. Borse, H.~Cai, N.~P. Pandey, F.~Yin, F.~Mayer \emph{et~al.}, ``{Real-Time, Accurate, and Consistent Video Semantic Segmentation via Unsupervised Adaptation and Cross-Unit Deployment on Mobile Device},'' in \emph{Proceedings of the IEEE/CVF Conference on Computer Vision and Pattern Recognition}, 2022, pp. 21\,431--21\,438.

\bibitem{sun2020mobilebert}
Z.~Sun, H.~Yu, X.~Song, R.~Liu, Y.~Yang, and D.~Zhou, ``{Mobilebert: A Compact Task-Agnostic Bert for Resource-Limited Devices},'' \emph{arXiv preprint arXiv:2004.02984}, 2020.

\bibitem{li2024snapfusion}
Y.~Li, H.~Wang, Q.~Jin, J.~Hu, P.~Chemerys, Y.~Fu, Y.~Wang, S.~Tulyakov, and J.~Ren, ``Snapfusion: Text-to-image diffusion model on mobile devices within two seconds,'' \emph{Advances in Neural Information Processing Systems}, vol.~36, 2024.

\bibitem{du2023exploring}
H.~Du, R.~Zhang, D.~Niyato, J.~Kang, Z.~Xiong, D.~I. Kim, H.~V. Poor \emph{et~al.}, ``{Exploring Collaborative Distributed Diffusion-Based AI-Generated Content (AIGC) in Wireless Networks},'' \emph{arXiv preprint arXiv:2304.03446}, 2023.

\bibitem{liu2024semantic}
G.~Liu, H.~Du, D.~Niyato, J.~Kang, Z.~Xiong, D.~I. Kim, and X.~Shen, ``Semantic communications for artificial intelligence generated content (aigc) toward effective content creation,'' \emph{IEEE Network}, 2024.

\bibitem{qin2021semantic}
Z.~Qin, X.~Tao, J.~Lu, and G.~Y. Li, ``{Semantic Communications: Principles and Challenges},'' \emph{arXiv preprint arXiv:2201.01389}, 2021.

\bibitem{xia2023wiservr}
L.~Xia, Y.~Sun, C.~Liang, D.~Feng, R.~Cheng, Y.~Yang, and M.~A. Imran, ``{WiserVR: Semantic Communication Enabled Wireless Virtual Reality Delivery},'' \emph{IEEE Wireless Communications}, vol.~30, no.~2, pp. 32--39, 2023.

\bibitem{liang2023vista}
C.~Liang, X.~Deng, Y.~Sun, R.~Cheng, L.~Xia, D.~Niyato, and M.~A. Imran, ``{VISTA: Video Transmission Over a Semantic Communication Approach},'' \emph{ICC 2023 - IEEE International Conference on Communication (ICC)}, 2023.

\bibitem{zhao2022background}
F.~Zhao, Y.~Sun, R.~Cheng, and M.~A. Imran, ``{Background Knowledge Aware Semantic Coding Model Selection},'' in \emph{2022 IEEE 22nd International Conference on Communication Technology (ICCT)}.\hskip 1em plus 0.5em minus 0.4em\relax IEEE, 2022, pp. 84--89.

\bibitem{sun2024s}
Y.~Sun, L.~Zhang, L.~Guo, J.~Li, D.~Niyato, and Y.~Fang, ``S-ran: Semantic-aware radio access networks,'' \emph{arXiv preprint arXiv:2407.11161}, 2024.

\bibitem{huang2021deep}
D.~Huang, X.~Tao, F.~Gao, and J.~Lu, ``{Deep Learning-Based Image Semantic Coding for Semantic Communications},'' in \emph{2021 IEEE Global Communications Conference (GLOBECOM)}.\hskip 1em plus 0.5em minus 0.4em\relax IEEE, 2021, pp. 1--6.

\bibitem{guler2018semantic}
B.~G{\"u}ler, A.~Yener, and A.~Swami, ``{The Semantic Communication Game},'' \emph{IEEE Transactions on Cognitive Communications and Networking}, vol.~4, no.~4, pp. 787--802, 2018.

\bibitem{liang2024generative}
C.~Liang, H.~Du, Y.~Sun, D.~Niyato, J.~Kang, D.~Zhao, and M.~A. Imran, ``Generative ai-driven semantic communication networks: Architecture, technologies and applications,'' \emph{IEEE Transactions on Cognitive Communications and Networking}, 2024.

\bibitem{pan2024sc}
H.~Pan, S.~Yang, T.-T. Chan, Z.~Wang, V.~C. Leung, and J.~Li, ``{SC-PNC: Semantic Communication-Empowered Physical-layer Network Coding},'' \emph{IEEE Transactions on Cognitive Communications and Networking}, 2024.

\bibitem{lin2023text}
Z.~Lin, Y.~Gong, Y.~Shen, T.~Wu, Z.~Fan, C.~Lin, N.~Duan, and W.~Chen, ``{Text Generation with Diffusion Language Models: A Pre-Training Approach with Continuous Paragraph Denoise},'' in \emph{International Conference on Machine Learning}.\hskip 1em plus 0.5em minus 0.4em\relax PMLR, 2023, pp. 21\,051--21\,064.

\bibitem{huang2022prodiff}
R.~Huang, Z.~Zhao, H.~Liu, J.~Liu, C.~Cui, and Y.~Ren, ``{Prodiff: Progressive Fast Diffusion Model for High-Quality Text-to-Speech},'' in \emph{Proceedings of the 30th ACM International Conference on Multimedia}, 2022, pp. 2595--2605.

\bibitem{rombach2022high}
R.~Rombach, A.~Blattmann, D.~Lorenz, P.~Esser, and B.~Ommer, ``{High-Resolution Image Synthesis with Latent Diffusion Models},'' in \emph{Proceedings of the IEEE/CVF conference on computer vision and pattern recognition}, 2022, pp. 10\,684--10\,695.

\bibitem{radford2021learning}
A.~Radford, J.~W. Kim, C.~Hallacy, A.~Ramesh, G.~Goh, S.~Agarwal, G.~Sastry, A.~Askell, P.~Mishkin, J.~Clark \emph{et~al.}, ``{Learning Transferable Visual Models from Natural Language Supervision},'' in \emph{International conference on machine learning}.\hskip 1em plus 0.5em minus 0.4em\relax PMLR, 2021, pp. 8748--8763.

\bibitem{kingma2013auto}
D.~P. Kingma and M.~Welling, ``{Auto-Encoding Variational Bayes},'' \emph{arXiv preprint arXiv:1312.6114}, 2013.

\bibitem{ho2020denoising}
J.~Ho, A.~Jain, and P.~Abbeel, ``{Denoising Diffusion Probabilistic Models},'' \emph{Advances in neural information processing systems}, vol.~33, pp. 6840--6851, 2020.

\bibitem{ronneberger2015u}
O.~Ronneberger, P.~Fischer, and T.~Brox, ``{U-net: Convolutional Networks for Biomedical Image Segmentation},'' in \emph{Medical Image Computing and Computer-Assisted Intervention--MICCAI 2015: 18th International Conference, Munich, Germany, October 5-9, 2015, Proceedings, Part III 18}.\hskip 1em plus 0.5em minus 0.4em\relax Springer, 2015, pp. 234--241.

\bibitem{jiang2022deep}
P.~Jiang, C.-K. Wen, S.~Jin, and G.~Y. Li, ``{Deep Source-Channel Coding for Sentence Semantic Transmission with HARQ},'' \emph{IEEE transactions on communications}, vol.~70, no.~8, pp. 5225--5240, 2022.

\bibitem{bansal2024cold}
A.~Bansal, E.~Borgnia, H.-M. Chu, J.~Li, H.~Kazemi, F.~Huang, M.~Goldblum, J.~Geiping, and T.~Goldstein, ``{Cold Diffusion: Inverting Arbitrary Image Transforms without Noise},'' \emph{Advances in Neural Information Processing Systems}, vol.~36, 2024.

\bibitem{kingma2016improved}
D.~P. Kingma, T.~Salimans, R.~Jozefowicz, X.~Chen, I.~Sutskever, and M.~Welling, ``{Improved Variational Inference with Inverse Autoregressive Flow},'' \emph{Advances in neural information processing systems}, vol.~29, 2016.

\bibitem{hosu2019effective}
V.~Hosu, B.~Goldlucke, and D.~Saupe, ``{Effective Aesthetics Prediction with Multi-Level Spatially Pooled Features},'' in \emph{proceedings of the IEEE/CVF conference on computer vision and pattern recognition}, 2019, pp. 9375--9383.

\bibitem{luo2021resource}
Q.~Luo, S.~Hu, C.~Li, G.~Li, and W.~Shi, ``{Resource Scheduling in Edge Computing: A Survey},'' \emph{IEEE Communications Surveys \& Tutorials}, vol.~23, no.~4, pp. 2131--2165, 2021.

\bibitem{croitoru2023diffusion}
F.-A. Croitoru, V.~Hondru, R.~T. Ionescu, and M.~Shah, ``Diffusion models in vision: A survey,'' \emph{IEEE Transactions on Pattern Analysis and Machine Intelligence}, vol.~45, no.~9, pp. 10\,850--10\,869, 2023.

\bibitem{wang2016dueling}
Z.~Wang, T.~Schaul, M.~Hessel, H.~Hasselt, M.~Lanctot, and N.~Freitas, ``{Dueling Network Architectures for Deep Reinforcement Learning},'' in \emph{International conference on machine learning}.\hskip 1em plus 0.5em minus 0.4em\relax PMLR, 2016, pp. 1995--2003.

\bibitem{hu2021lora}
E.~J. Hu, Y.~Shen, P.~Wallis, Z.~Allen-Zhu, Y.~Li, S.~Wang, L.~Wang, and W.~Chen, ``Lora: Low-rank adaptation of large language models,'' \emph{arXiv preprint arXiv:2106.09685}, 2021.

\end{thebibliography}

\begin{IEEEbiography}[{\includegraphics[width=1in,height=1.25in,clip,keepaspectratio]{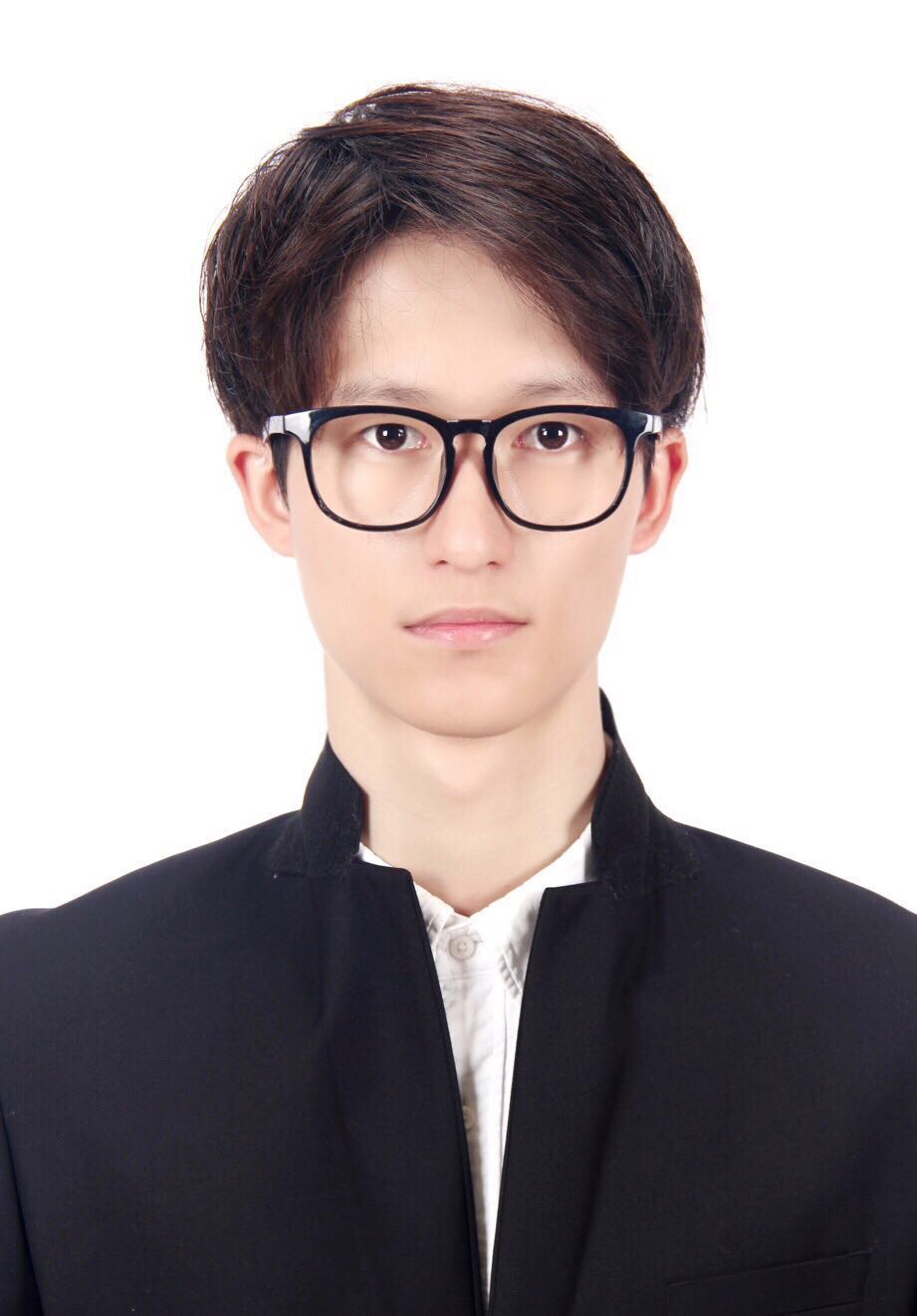}}]{Runze Cheng}
    (Member, IEEE) received the Ph.D. degree in Electrical and Electronic Engineering from the James Watt School of Engineering, University of Glasgow, U.K. in 2023. He is currently a Postdoctoral Research Associate with the Communications, Sensing, and Imaging Research Group, University of Glasgow, Glasgow, U.K. His research interests include intelligent resource management, semantic communication, distributed machine learning, and space-air-ground integrated networks.  
\end{IEEEbiography}

\begin{IEEEbiography}[{\includegraphics[width=1in,height=1.25in,clip,keepaspectratio]{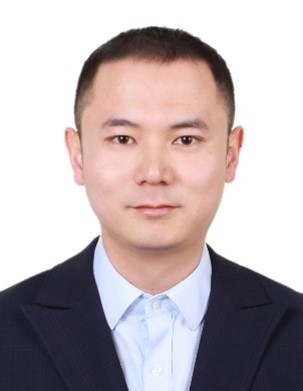}}]{Yao Sun}
    (Senior Member, IEEE) is currently a Lecturer with James Watt School of Engineering, the University of Glasgow, Glasgow, UK. Dr. Sun has extensive research experience in wireless communication area. He has won the IEEE IoT Journal Best Paper Award 2022, and IEEE Communication Society of TAOS Best Paper Award in 2019 ICC. His research interests include intelligent wireless networking, SemCom and wireless blockchain system. 
\end{IEEEbiography}

\begin{IEEEbiography}[{\includegraphics[width=1in,height=1.25in,clip,keepaspectratio]{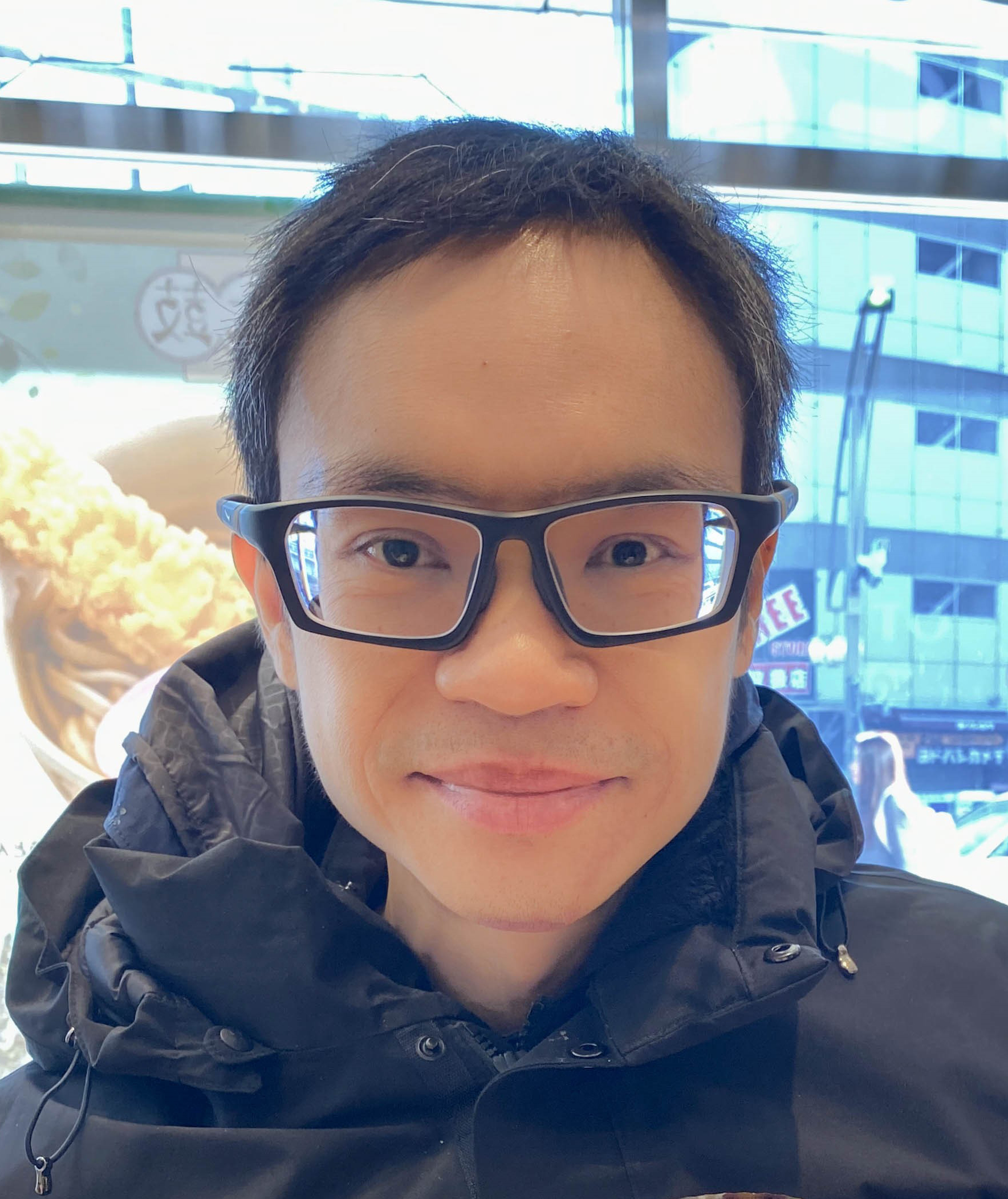}}]{Dusit Niyato}
    (Fellow, IEEE) is a professor in the College of Computing and Data Science, at Nanyang Technological University, Singapore. He received B.Eng. from King Mongkuts Institute of Technology Ladkrabang (KMITL), Thailand and Ph.D. in Electrical and Computer Engineering from the University of Manitoba, Canada. His research interests are in the areas of mobile generative AI, edge intelligence, decentralized machine learning, and incentive mechanism design.
\end{IEEEbiography}

\begin{IEEEbiography}[{\includegraphics[width=1in,height=1.25in,clip,keepaspectratio]{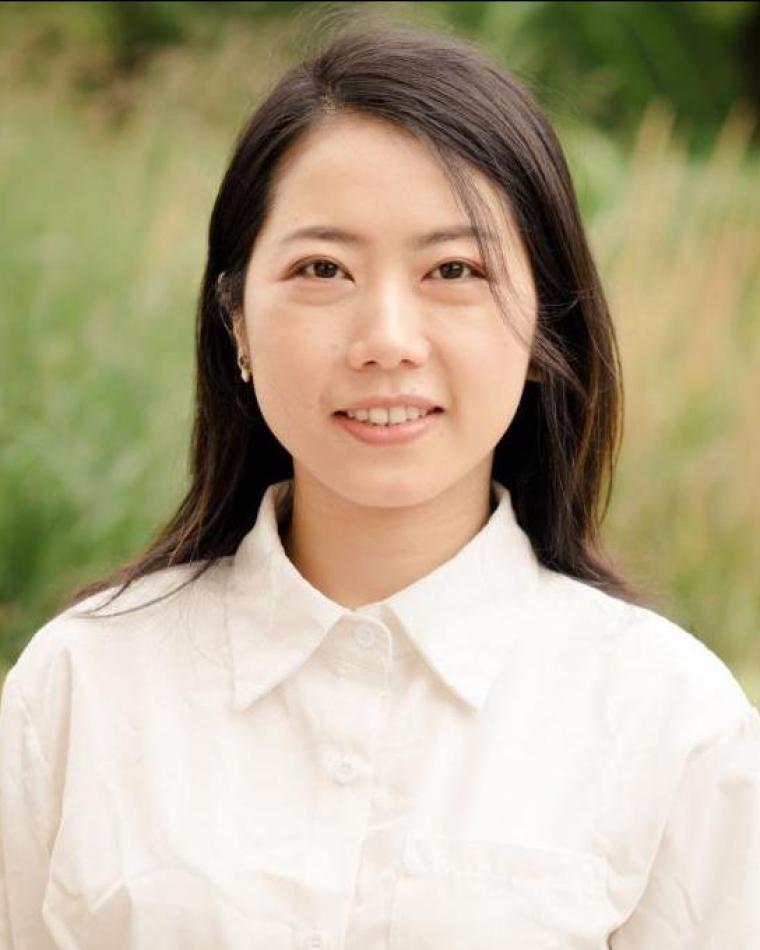}}]{Lan Zhang}
    (Member, IEEE) received the BE and MS degrees from the University of Electronic Science and Technology of China in 2013 and 2016, respectively, and the PhD degree from the University of Florida in 2020. She has been a tenure-track assistant professor with the Department of Electrical and Computer Engineering at Clemson University since 2024. Before that, she was an assistant professor with the Department of Electrical and Computer Engineering at Michigan Technological University from 2020 to 2023. Her research interests include wireless communications, distributed machine learning, and cybersecurity for various Internet-of-Things applications.
\end{IEEEbiography}

\begin{IEEEbiography} [{\includegraphics[width=1in,height=1.25in,clip,keepaspectratio]{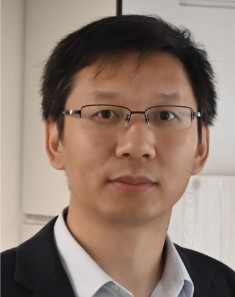}}]{Lei Zhang}  
    (Senior Member, IEEE) is a Professor of Trustworthy Systems with the University of Glasgow, Glasgow, U.K. He has academia and industry combined research experience on wireless communications and networks, and distributed systems for the Internet of Things, blockchain, and autonomous systems. His 20 patents are granted/filed in more than 30 countries/regions. Dr. Zhang received the IEEE ComSoc TAOS Technical Committee Best Paper Award in 2019, the IEEE Internet of Things Journal Best Paper Award in 2022, Digital Communications and Networks Journal Best Paper Award 2023 in addition to several best paper awards in IEEE conferences. He is the Founding Chair of the IEEE Special Interest Group on Wireless Blockchain Networks in the IEEE Cognitive Networks Technical Committee (TCCN). He is an Associate Editor of IEEE Internet of Things Journal, IEEE Wireless Communications Letters, IEEE Transactions on Network Science and Engineering, and Digital Communications and Networks.
\end{IEEEbiography}

\begin{IEEEbiography} [{\includegraphics[width=1in,height=1.25in,clip,keepaspectratio]{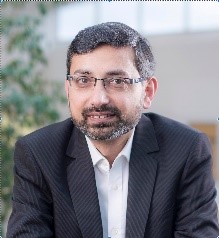}}]{Muhammad Ali Imran}  
    (Fellow, IEEE) is a Professor of Wireless Communication Systems and Dean of Graduate Studies in College of Science and Engineering. His research interests include self-organized networks, wireless networked control systems, and the wireless sensor systems. He heads the Communications, Sensing and Imaging CSI Hub, University of Glasgow, Glasgow, U.K. He is also an Affiliate Professor with The University of Oklahoma, Norman, OK, USA, and a Visiting Professor with the 5G Innovation Centre, University of Surrey, Guildford, U.K. He has more than 20 years of combined academic and industry experience with several leading roles in multimillion pounds funded projects. His research interests include self-organized networks, wireless networked control systems, and the wireless sensor systems.
\end{IEEEbiography}

\end{document}